\documentclass[twocolumn,superscriptaddress,amsmath,amssymb,prx,floatfix]{revtex4-2}

\usepackage{xr}
\usepackage{hyperref}
\hypersetup{
    colorlinks=true,
    linkcolor=blue,
    filecolor=magenta,      
    urlcolor=cyan,
    pdfpagemode=FullScreen,
    }

\usepackage{graphicx}
\usepackage{dcolumn}
\usepackage{bm}

\usepackage{threeparttable}

\usepackage[utf8]{inputenc}
\usepackage[T1]{fontenc}
\usepackage{mathptmx}
\usepackage[dvipsnames]{xcolor}
\usepackage{geometry}
\usepackage[explicit]{titlesec}

\usepackage{xcolor}
\usepackage[draft]{changes}
\definechangesauthor[color=red, name={Paulo Santos}]{PVS}
\definechangesauthor[color=cyan, name={Alexander Kuznetsov}]{AK}

\newcommand{\hBlue}[1]{{\color{black}{#1}}}

\usepackage{verbatim}
\setauthormarkup{}

\RequirePackage[normalem]{ulem} 
\RequirePackage{color}\definecolor{RED}{rgb}{1,0,0}\definecolor{BLUE}{rgb}{0,0,1} 
\providecommand{\DIFaddbegin}{} 
\providecommand{\DIFaddend}{} 
\providecommand{\DIFdelbegin}{} 
\providecommand{\DIFdelend}{} 
\providecommand{\DIFaddbeginFL}{} 
\providecommand{\DIFaddendFL}{} 
\providecommand{\DIFdelbeginFL}{} 
\providecommand{\DIFdelendFL}{} 
\newcommand{\DIFscaledelfig}{0.5}
\RequirePackage{settobox} 
\RequirePackage{letltxmacro} 
\newsavebox{\DIFdelgraphicsbox} 
\newlength{\DIFdelgraphicswidth} 
\newlength{\DIFdelgraphicsheight} 
\LetLtxMacro{\DIFOincludegraphics}{\includegraphics} 
\newcommand{\DIFaddincludegraphics}[2][]{{\color{blue}\fbox{\DIFOincludegraphics[#1]{#2}}}} 
\newcommand{\DIFdelincludegraphics}[2][]{
\sbox{\DIFdelgraphicsbox}{\DIFOincludegraphics[#1]{#2}}
\settoboxwidth{\DIFdelgraphicswidth}{\DIFdelgraphicsbox} 
\settoboxtotalheight{\DIFdelgraphicsheight}{\DIFdelgraphicsbox} 
\scalebox{\DIFscaledelfig}{
\parbox[b]{\DIFdelgraphicswidth}{\usebox{\DIFdelgraphicsbox}\\[-\baselineskip] \rule{\DIFdelgraphicswidth}{0em}}\llap{\resizebox{\DIFdelgraphicswidth}{\DIFdelgraphicsheight}{
\setlength{\unitlength}{\DIFdelgraphicswidth}
\begin{picture}(1,1)
\thicklines\linethickness{2pt} 
{\color[rgb]{1,0,0}\put(0,0){\framebox(1,1){}}}
{\color[rgb]{1,0,0}\put(0,0){\line( 1,1){1}}}
{\color[rgb]{1,0,0}\put(0,1){\line(1,-1){1}}}
\end{picture}
}\hspace*{3pt}}} 
} 
\LetLtxMacro{\DIFOaddbegin}{\DIFaddbegin} 
\LetLtxMacro{\DIFOaddend}{\DIFaddend} 
\LetLtxMacro{\DIFOdelbegin}{\DIFdelbegin} 
\LetLtxMacro{\DIFOdelend}{\DIFdelend} 
\DeclareRobustCommand{\DIFaddbegin}{\DIFOaddbegin \let\includegraphics\DIFaddincludegraphics} 
\DeclareRobustCommand{\DIFaddend}{\DIFOaddend \let\includegraphics\DIFOincludegraphics} 
\DeclareRobustCommand{\DIFdelbegin}{\DIFOdelbegin \let\includegraphics\DIFdelincludegraphics} 
\DeclareRobustCommand{\DIFdelend}{\DIFOaddend \let\includegraphics\DIFOincludegraphics} 
\LetLtxMacro{\DIFOaddbeginFL}{\DIFaddbeginFL} 
\LetLtxMacro{\DIFOaddendFL}{\DIFaddendFL} 
\LetLtxMacro{\DIFOdelbeginFL}{\DIFdelbeginFL} 
\LetLtxMacro{\DIFOdelendFL}{\DIFdelendFL} 
\DeclareRobustCommand{\DIFaddbeginFL}{\DIFOaddbeginFL \let\includegraphics\DIFaddincludegraphics} 
\DeclareRobustCommand{\DIFaddendFL}{\DIFOaddendFL \let\includegraphics\DIFOincludegraphics} 
\DeclareRobustCommand{\DIFdelbeginFL}{\DIFOdelbeginFL \let\includegraphics\DIFdelincludegraphics} 
\DeclareRobustCommand{\DIFdelendFL}{\DIFOaddendFL \let\includegraphics\DIFOincludegraphics} 
\makeatletter 
\let\sout@orig\sout 
\renewcommand{\sout}[1]{\ifmmode\text{\sout@orig{\ensuremath{#1}}}\else\sout@orig{#1}\fi} 
\makeatother 
\RequirePackage{listings} 
\RequirePackage{color} 
\lstdefinelanguage{DIFcode}{ 
  moredelim=[il][\color{red}\sout]{\%DIF\ <\ }, 
  moredelim=[il][\color{blue}\uwave]{\%DIF\ >\ } 
} 
\lstdefinestyle{DIFverbatimstyle}{ 
	language=DIFcode, 
	basicstyle=\ttfamily, 
	columns=fullflexible, 
	keepspaces=true 
} 
\lstnewenvironment{DIFverbatim}{\lstset{style=DIFverbatimstyle}}{} 
\lstnewenvironment{DIFverbatim*}{\lstset{style=DIFverbatimstyle,showspaces=true}}{} 
\begin{document}

\title{Deterministic single-photon source with on-chip 5.6 GHz acoustic clock}

\author{Alexander S. Kuznetsov}
\email[corresponding author: ]{kuznetsov@pdi-berlin.de}
\affiliation{Paul-Drude-Institut f{\"u}r Festk{\"o}rperelektronik, Leibniz-Institut im Forschungsverbund Berlin e.V., Hausvogteiplatz 5-7, 10117 Berlin, Germany}

\author{Meysam Saeedi}
\affiliation{Paul-Drude-Institut f{\"u}r Festk{\"o}rperelektronik, Leibniz-Institut im Forschungsverbund Berlin e.V., Hausvogteiplatz 5-7, 10117 Berlin, Germany}

\author{Zixuan Wang}
\affiliation{Department of Physics, University of Colorado, Boulder, Colorado 80309, USA}
\affiliation{National Institute of Standards and Technology, Boulder, Colorado 80305, USA}

\author{Kevin L. Silverman}
\affiliation{National Institute of Standards and Technology, Boulder, Colorado 80305, USA}

\author{Klaus Biermann}
\affiliation{Paul-Drude-Institut f{\"u}r Festk{\"o}rperelektronik, Leibniz-Institut im Forschungsverbund Berlin e.V., Hausvogteiplatz 5-7, 10117 Berlin, Germany}

\date{\today}
\begin{abstract}
Scalable solid-state single-photon sources with triggered single-photon emission rates exceeding a few GHz would aid in the wide technological adoption of photonic quantum technologies. We demonstrate triggering of a quantum dot (QD) single photon emission using dynamic Purcell effect induced at a frequency of several GHz by acoustic strain. To this end, InAs QDs are integrated in a hybrid photon-phonon patterned microcavity \hBlue{with engineered optical and acoustic phonon degrees of freedom.} The density of optical states is tailored by the lateral confinement of photons in $\mu$m-sized traps defined lithographically in the microcavity spacer. The single-photon character of the emission from a QD in a trap is confirmed by measuring single-photon statistics. We demonstrate modulation of the QD transition energy in a trap with a frequency up to 14~GHz by monochromatic longitudinal bulk acoustic phonons generated by piezoelectric transducers. For the modulation frequency of 5.6~GHz, corresponding to the acoustic mode of the microcavity, the QD energy is periodically shifted through a spectrally narrow confined photonic mode leading to an appreciable enhancement of the QD emission due to the dynamic Purcell effect. 
    \hBlue{We numerically explored time-evolution of the QD-cavity system under modulation and identified conditions that could enable single-photon rates exceeding 5~GHz.}
The platform thus enables the implementation of scalable III-V-based single-photon sources with on-chip tunable acoustic clocks with frequencies that can exceed several GHz under continuous wave optical excitation.

\end{abstract}

\maketitle 




\section{Introduction}
\label{Introduction}

Since the first experimental observation of anti-bunching heralding single-photon (SP) emission~\cite{kimble1977Phys.Rev.Lett.}, single-photon sources (SPSs) have been identified as a key component of optical quantum technologies~\cite{senellart2017Nat.Nanotechnol.}. Among a great variety of SPSs, deterministic ones based on semiconductor quantum dots (QDs)~\cite{michler2000Science} arguably have the highest relevance for high-performance applications~\cite{liu2025Appl.Phys.Rev.}. 
    \hBlue{High SP emission rates are crucial for optical quantum computing, quantum repeaters, quantum key distribution, quantum sensing and imaging, to name a few~\cite{couteau2023NatRevPhys}.}
%

Under continuous wave optical excitation, solid-state QDs can produce SPs with spontaneous emission (SE) rates up to $\gamma_{SE}^{cav} \approx 80$~GHz~\cite{hoang2016NanoLett.}, owing to the shortened radiative lifetimes achieved by photonic engineering. However, the best commercial deterministic SPS are limited to triggered SP rates into first lens below a GHz~\cite{georgieva2024Metrologia}. \hBlue{This discrepancy underscores potential orders-of-magnitude improvement to the efficiency of SPSs.}

A well-known strategy towards high-rate SP emission is to combine GHz-rate excitation with Purcell enhancement of the QD radiative rate achieved by engineering of the photonic density of states around a QD~\cite{Purcell1946Phys.Rev.}. As the result, the radiative lifetime $\tau_{SE}^{cav} = 1/\gamma_{SE}^{cav}$ in cavity is shortened by the Purcell factor ($F_{P}$) compared to the isotropic one $\tau_{SE}^{cav} = \tau_{SE}^{iso}/F_{P}$.
Such enhancement has been demonstrated in photonic structures and cavities based on nanowires~\cite{claudon2010Nat.Photonics}, micropillars~\cite{gazzano2013Nat.Commun.}, \hBlue{etched-and-overgrown mesas~\cite{engel2023Appl.Phys.Lett.}}, photonic crystals~\cite{madsen2014Phys.Rev.B}, circular Bragg gratings~\cite{sapienza2015Nat.Commun., abudayyeh2021ACSNano}, open resonators~\cite{tomm2021Nat.Nanotechnol.} and broadband cylindrical nanocavities~\cite{chellu2024ArXiv}. 

    \hBlue{The demonstrated large values of $\gamma_{SE}^{cav} \approx 30$~GHz, can, in principle, facilitate triggered high-purity SP emission with rates $\gamma_{SP} \approx \gamma_{SE}^{cav}/n$ exceeding a few GHz, where integer $n \approx 5$~\cite{rickert2025ACSPhotonics}. Purcell enhancement factors $F_{P}$ reaching 70 have been predicted for etched-and-overgrown cavities~\cite{sultanivala2025J.Phys.Photonics}. In combination with precise, deterministic fabrication of photonic cavities~\cite{rodt2021NanoExpress} with $F_{P} > 20$ embedding fast SP emitters, e.g., GaAs QDs with isotropic $\tau_{SE}^{iso} < 200$~ps~\cite{hours2005Phys.Rev.B}, $\gamma_{SE}^{cav} \approx 100$~GHz could be anticipated, potentially, leading to scalable on-chip SPSs with giant $\gamma_{SP} > 20$~GHz, provided an equally fast and efficient on-chip triggering mechanism is realized.}

    \hBlue{Laboratory quantum light sources have reached trigger rates ($F_{trig}$) up to 3 GHz using optical~\cite{rickert2025ACSPhotonics,anderson2020Phys.Rev.Applied, poortvliet2025Phys.Rev.Appl.} and electrical pulses~\cite{hargart2013Appl.Phys.Lett.,shooter2020Opt.Express,zhang2015Nat.Commun.,muller2020Phys.Rev.Research}, as well as hybrid excitation~\cite{schlehahn2016Appl.Phys.Lett.}. However, these solutions present challenges, e.g., high-$F_{trig}$ optical approaches require complex external electro-optic modulators~\cite{anderson2020Phys.Rev.Applied,poortvliet2025Phys.Rev.Appl.} or bulky frequency multipliers~\cite{rickert2025ACSPhotonics}, whereas electrical triggering reduces SP purity and indistinguishability at higher $F_{trig}$ due to its non-resonant  character~\cite{hargart2013Appl.Phys.Lett.,muller2020Phys.Rev.Research}. 
    }

    \hBlue{Thus, an important milestone towards ultra-fast SPSs is to realize GHz-fast triggering mechanism that is on-chip and scalable, can be integrated with high quality-factor (Q) photonic structures and does not negatively affect properties of the quantum light.}

\begin{figure*}[!t]
	\centering
		\includegraphics[width=1\textwidth, keepaspectratio=true]{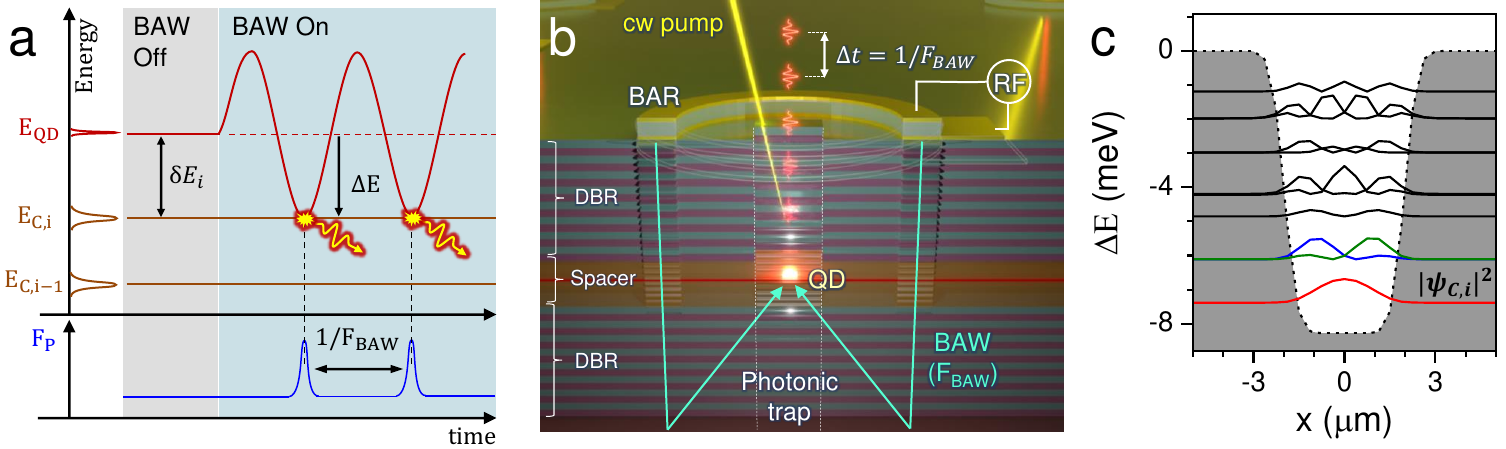}
		\caption{
            {\bf Dynamic Purcell effect under acoustic modulation.}
            {\bf a} Schematic energy diagram of the system. QD energy is $E_\mathrm{QD}$. The energy of an $i^\mathrm{th}$ confined cavity mode is $E_\mathrm{C,i}$. The spectral width of the resonances is indicated by Gaussian peaks. Purcell enhancement factor is $F_\mathrm{P}$. The initial cavity-QD detuning (i.e., without the modulation -- RF Off) is $\Delta E$. The QD energy is modulated by a BAW, RF On, which results in a periodic resonance with the cavity mode. For simplicity, we assume that the energy modulation amplitude is equal to the static QD-cavity detuning. At resonance there is a large Purcell enhancement leading to the triggered emission of single photons (wiggly lines). 
            {\bf b} A sketch of the MC platform to realize the modulation of panel $a$. An InAs QD is located within a spacer of an AlGaAs MC at the center of a lateral photonic trap, defined by the thicker part of the spacer. A bulk acoustic resonator, BAR, converts an applied radio-frequency signal into a BAW of GHz frequency ($F_\mathrm{BAW}$). The BAW propagates towards the center of the trap and modulates the QD via the deformation potential coupling. The device converts a CW pump into a stream of single photons with a rate equal to $F_\mathrm{BAW}$.
            {\bf c} An exemplary calculated spectrum of a $4 \times 4 ~ \mu \mathrm{m}^2$ photonic trap. The dashed line designates the confining potential, defined by the variation of the spacer thickness. The solid lines are spatial profiles of the squared confined wavefunctions.   
            }
	\label{Fig1}
\end{figure*}

    \hBlue{
    A potential solution is based on the fast modulation of the resonance condition between the QD transition energy and photonic resonance in a QD-cavity system, leading to a dynamical Purcell effect. In high-Q cavities, this would result in the generation of short SP pulses, as illustrated in Fig.~\ref{Fig1}a for the case of the fast crossing between the QD resonance and a photonic mode. So far, this concept has been demonstrated with \hBlue{MHz-to-GHz frequency Stark tuning~\cite{pagliano2014NatCommun}}, dynamical strain tuning by 800~MHz surface acoustic waves (SAWs)~\cite{weiss2016Appl.Phys.Lett.} and fast refractive index changes induced by ps-scale optical pulses with MHz repetition rates~\cite{peinke2021LightSci.Appl.}.
    }

    \hBlue{Among the dynamical approaches mentioned above, modulation of photonic structures using dynamical acoustic strain fields offers a pathway towards coherent modulation using compact on-chip piezoelectric transducers with very high-rates $F_{trig} > 1$~GHz~\cite{fuhrmann2011Nat.Photonics}. Acoustic strain can couple to the cavity optical mode via the variation of the refractive index and geometric deformation, and to excitonic transitions via the deformation potential (DP) mechanism and piezoelectric fields~\cite{delima2005Rep.ProgPhys}.} 
Acoustic fields have enabled modulation of transition energies~\cite{akimov2006Phys.Rev.Lett.}, transport of excitons~\cite{lazic2010PhysELow-Dimens.SystNanostructures} and coherent control of QDs~\cite{buhler2022Nat.Commun., decrescent2024Optica, Zhan2026Phys.Rev.Lett.a}. These effects have been used to demonstrate the unconventional triggering of SP emission from bare QDs~\cite{couto2009Nat.Photonics}, QDs in photonic crystals~\cite{weiss2016Appl.Phys.Lett.} and micropillars~\cite{villa2017Appl.Phys.Lett.} with sub-GHz frequencies. 
    \hBlue{Other original approaches use the ability of SAWs to carry electrons and excitons to produce GHz-clocked SPs from n-i-p junctions and defects~\cite{hsiao2020NatCommun,yuan2021ACSPhotonics}.} 
The fastest acoustic clock reported so far has been demonstrated for bare color centers remotely pumped by a 3.5~GHz SAW~\cite{yuan2021ACSPhotonics}. Complementary, static strain tuning has been used to reach very high resonant Purcell enhancements~\cite{mudi2024Appl.Phys.Lett.,yang2024LightSci.Appl.}.

We have demonstrated that the exciton transition energy of a quantum well exciton embedded in an optomechanical microcavity (MC) can be modulated using monochromatic longitudinal bulk acoustic waves (BAWs) at frequencies that reach 20~GHz~\cite{kuznetsov2021Phys.Rev.X} and very large amplitudes of 50~meV~\cite{sesin2023Phys.Rev.Res.} generated using compact piezoelectric acoustic transducers~\cite{machado2019Phys.Rev.Appl.}. This large energy modulation results in periodic crossings at GHz frequency with multiple spectrally narrow confined optical modes with energy rates that can reach $dE/dt \sim 1$~meV/ps~\cite{kuznetsov202a6Nat.Photon.}. 

    \hBlue{In this work, we embedded an InAs QD in a photonic trap created in the spacer of a hybrid photon-phonon AlGaAs MC. Then, we demonstrated acoustic modulation of the QD energy up to 14 GHz using dynamical strain of a standing BAW generated using a piezolectric transducer. For the frequency of acoustic cavity mode at 5.6 GHz we achieved dynamical crossing between QD and cavity modes. Under these conditions we observed suppression and enhancement of the QD emission, which we are known signatures of dynamical Purcell effect~\cite{pagliano2014NatCommun, weiss2016Appl.Phys.Lett.}. Thus, we demonstrated that BAW acts as on-chip trigger of SP emission. Our numerical modeling of time-evolution of the QD-cavity system shows that triggered high-purity SP emission with rates exceeding $\gamma_{SP} > 5$~GHz can be reached in realistic structures.}


\section{Results}
\label{Results}

\begin{figure*}[t]
	\centering
		\includegraphics[width=1\textwidth, keepaspectratio=true]{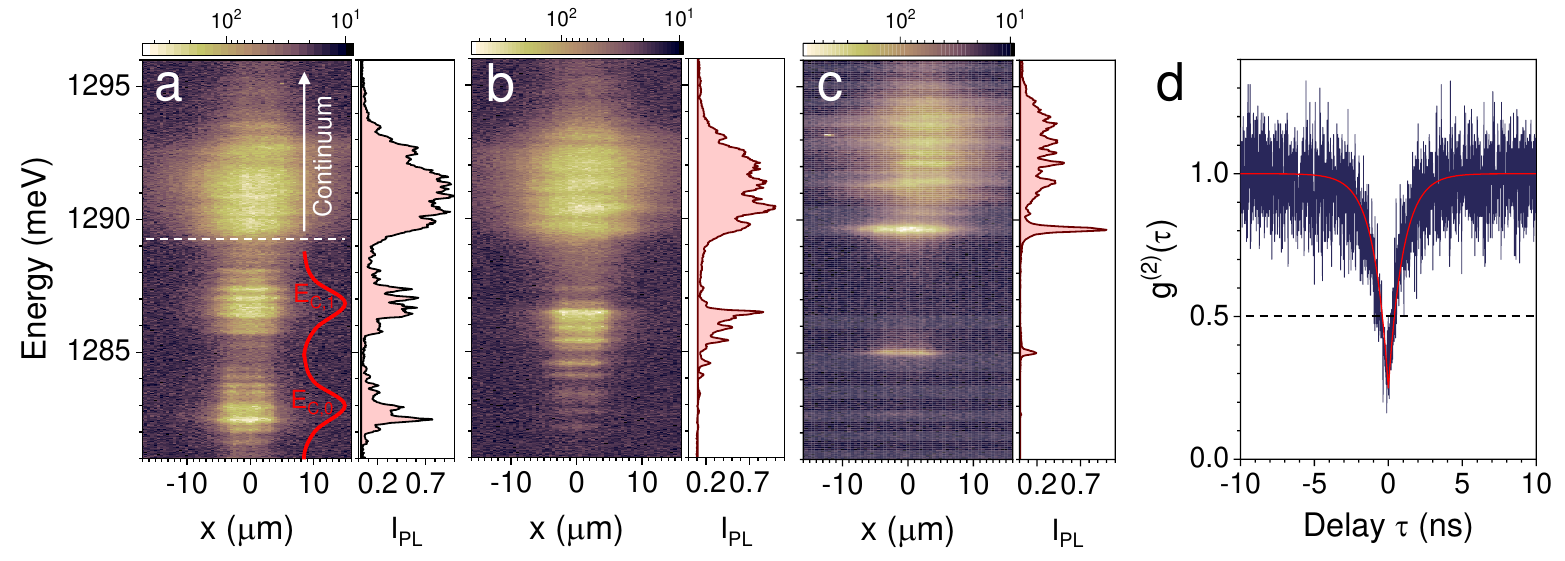}
		\caption{
            {\bf Quantum dots in photonic traps.}
            {\bf a} Spatially and spectrally resolved PL of a $2 \times 2 ~ \mu \mathrm{m}^2$ trap in a high-density of QDs part of the sample under weak non-resonant excitation. The panel on the right is the spatially integrated spectrum. Confined photonic modes are designated $E_\mathrm{C,0}$ and $E_\mathrm{C,1}$.
            {\bf b, c} PL of a $1 \times 1 ~ \mu \mathrm{m}^2$ trap in the high-density and low-density part of the sample, respectively. The excitation conditions are similar to those of panel $a$. The color bars encode the PL intensity in arbitrary units.
            {\bf d} A second-order auto-correlation measurement as a function of delay ($g^{(2)}(\tau)$) of a single PL peak from a trap in a low-density region. \hBlue{The red curve is the fit of the dip edges with mono-exponential functions.}
            }
	\label{Fig2}
\end{figure*}

To realize the dynamic Purcell effect of Fig.~\ref{Fig1}a, we designed and fabricated a structure, schematically shown in Fig.~\ref{Fig1}b. The device comprises an AlGaAs MC. Its spacer contains a single layer of InAs self-assembled QDs. The MC details are presented in Methods. It is designed such that the spacer co-localizes (in the growth direction) a 1.3~eV photonic mode ($Q_{C} \approx 20000$) and a 6~GHz ($Q_{acc} \approx 70$) acoustic phonon mode. The calculated depth profiles and reflectivity spectra of the optical and acoustic fields are presented in the Supplementary Information SI-Sec.1. The position of the QD layer is chosen to maximize the coupling to the optical and acoustic fields (i.e., close to the anti-nodes of the electric and strain fields). After fabricating the lower DBR and the spacer with QDs, the spacer of the MC is etched producing a few-nm-high and a few-$\mu$m-wide mesas. The etching shifts the cavity mode by approximately 10~meV (SI-Sec.2). The sample is then overgrown with the upper DBR. The mesas act as photonic traps, which create zero-dimensional photonic states. The lateral size and the shape of a trap defines the spectrum of the confined modes, e.g., Fig.~\ref{Fig1}c. 
    \hBlue{The calculated spectrum presents a general structure of confined photonic modes for realistic cavity and trap parameters using the calculation procedure established in Ref.~\cite{kuznetsov2018Phys.Rev.B}. Simulations for other relevant trap sizes are presented in SI-Sec.9.} 
The density of QDs, which ranges from tens~$\mu \mathrm{m}^{-2}$ to $1 ~ \mu \mathrm{m}^{-2}$ and the large number of traps ensure high likelihood of their spatial overlap. Bulk acoustic wave resonators (BARs) are fabricated on the upper DBR around the traps. The active area of a BAR has an aperture for the optical access. As will be discussed later, a monochromatic GHz strain can be injected into the trap by exciting the BAR with a microwave signal.

We first discuss emission of QDs in traps in the absence of acoustic modulation. Experiments have been carried out at 5~K as described in Methods. Figure~\ref{Fig2}a shows a spatially and spectrally resolved photoluminescence (PL) map of a $2 \times 2 ~ \mu \mathrm{m}^2$ trap in the high QD density sample region under weak continuous wave non-resonant excitation ($\lambda_\mathrm{laser} \approx 804$~nm, $P_\mathrm{laser} \approx 30 ~ \mu$W). We identify three emission bands. The band above 1289~meV is a continuum of un-confined states with a large spatial extent. Below 1289~meV there are two modes (designated $E_\mathrm{C,i}$, where $i = \{0,1\}$) localized in energy and space, which are the confined modes of the trap. The continuum, corresponds to the parabolic cavity dispersion in the spatially extended etched region, while confined photonic states have flat in-plane momentum (SI-Sec.4). The spatially integrated PL spectrum shown in right side of panel of Fig.~\ref{Fig2}a, reveals that the emission consists of overlapping narrow peaks, which correspond to different QDs (and possibly to various excitonic complexes). In essence, the confined cavity modes act as a filter for the emission of QDs. From the momentum-resolved (SI-Sec.4) and the excitation power dependent PL measurements (SI-Sec.5), we extract the linewidth and hence the quality factor of both the extended region and the confined modes, which is about $Q = 1500 \pm 100$ and several times smaller than the typical values of 5000--10000 for this type of MCs~\cite{kuznetsov2023Nat.Commun.}. 
    \hBlue{The likely origin of the problem is the surface roughness (see SI-Sec. 8), which sometimes appears during the MBE-overgrowth. Therefore, presently low-Q is not a fundamental issue, but a matter of improving fabrication (e.g., optimization of overgrowth conditions).}

\begin{figure*}[!t]
	\centering
		\includegraphics[width=1\textwidth, keepaspectratio=true]{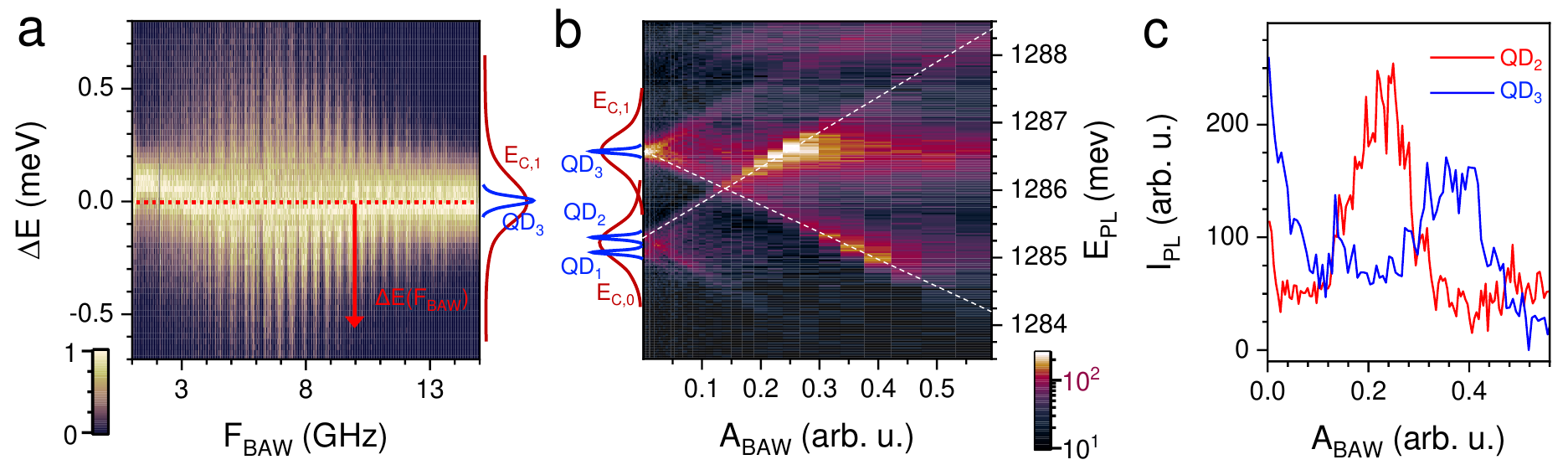}
		\caption{
            {\bf GHz acoustic modulation of QDs in photonic traps.}
            {\bf a} Spectral dependence of a single narrow line resonant to a confined level of a $4 \times 4 ~ \mu m^2$ trap on BAW frequency ($F_\mathrm{BAW}$) for a fixed amplitude ($A_\mathrm{BAW} = 0.56$). The energy is referenced to 1286.7~meV. $\Delta E (F_{BAW})$ designates the energy modulation amplitude for a given value of $F_\mathrm{BAW}$. For clarity, the spectrum for each frequency was normalized to its maximum.
            {\bf b} Dependence of the lower confined levels of the same trap on $A_\mathrm{BAW}$ for $F_\mathrm{BAW} = 5.6$~GHz. To improve the QDs energy shifts, the data was filtered using a bandpass filter. The original data and the explanation of the filtering procedure is described in SI-Sec.7. The color bars in panels $a$ and $b$ encode the PL intensity in arbitrary units.
            {\bf c} Intensity profiles taken along the dashed lines in the panel $b$ for the resonances designated as $QD_2$ and $QD_3$.
            }
	\label{Fig3}
\end{figure*}

In the case of a nearby $1 \times 1 ~ \mu \mathrm{m}^2$ trap, there is only one confined mode below the continuum, cf. Fig.~\ref{Fig2}b. The dependence of the trap spectrum on the lateral trap dimension is presented in SI-Sec.3. 
    \hBlue{We note a generally good agreement between the measured spectra and the calculated ones, see SI-Sec.~9.}
Figure~\ref{Fig2}c presents a PL map of another $1 \times 1 ~ \mu \mathrm{m}^2$ trap located in a region of the sample with low QD density. In striking contrast to the PL of Fig.~\ref{Fig2}b, only two narrow resonances can be observed below the continuum. This demonstrates that our approach can produce traps with few or single QDs. In order to investigate photon statistics, we measured the second-order auto-correlation function ($g^{(2)}(\tau)$) of a single emission line in a trap. A $g^{(2)}(\tau)$ trace under non-resonant excitation shows a clear dip around delay time $\tau = 0$, cf. Fig.~\ref{Fig2}d. 
    \hBlue{We fitted the $g^{(2)}(\tau)$ dip edges with mono-exponential functions (solid red curves in Fig.~\ref{Fig2}d) with decay time $\tau_{g^{(2)}} = 1$~ns and $g^{(2)}(0) = 0.25$. The latter value indicates a clear anti-bunched emission, demonstrating SP character of PL from a QD in a trap. The relatively high value of $g^{(2)}(0)$ is due to the emission from spectrally close resonances, see SI-Sec.~12.}

We now proceed to discuss the effect of the GHz bulk acoustic wave on the QD emission in the MC. Firstly, we recall that the MC is designed to predominantly couple BAW strain to the QD exciton via the DP mechanism characterized by the coupling strength on the order of $20 \times 10^6$~MHz, which is several orders of magnitude larger than the photoelastic coupling~\cite{sesin2023Phys.Rev.Res.}. It is this property that allows us to realize the modulation sketched in Fig.~\ref{Fig1}a. 

For the modulation experiment we focused on a $4 \times 4 ~ \mu \mathrm{m}^2$ trap in the low QD density region of the sample, which, unlike the traps discussed above, is located within the aperture of a BAR optimized to excite a 6~GHz BAW. The large generation bandwidth of the BAR allows to generate monochromatic BAWs tunable in the 2--15~GHz range. The electrical characteristic of BARs are presented in SI-Sec.~6. 

Figure~\ref{Fig3}a shows the dependence of the normalized PL of a QD in resonance with a confined optical mode of the trap as a function of the BAW frequency ($F_\mathrm{BAW}$) for a fixed BAW amplitude. This time-integrated image clearly shows characteristic signatures of the QD energy modulation, namely, the spectral broadening of the PL line with a frequency-dependent amplitude $\Delta E (F_{BAW})$. This symmetric broadening corresponds to the harmonic modulation of the QD resonance at $F_\mathrm{BAW}$. The largest $\Delta E (F_{BAW})$ values are achieved around 6~GHz, the intended frequency of the MC. Remarkably, the modulation with an amplitude greater than the linewidth is observed up to a maximum frequency of 14~GHz, which is limited by the bandwidth of the BAW transducer.

We now turn to the experimental demonstration of the dynamic Purcell effect. Figure~\ref{Fig3}b shows the dependence of the trap spectrum on the BAW amplitude ($A_\mathrm{BAW}$), which is proportional to the microwave (RF) power applied to the transducer, for $F_\mathrm{BAW} = 5.6$~GHz. For $A_\mathrm{BAW} = 0$, two narrow resonances (designated $QD_1$ and $QD_2$) overlap with the photonic ground state ($E_\mathrm{C,0}$), and a single bright resonance ($QD_3$) is resonant with $E_\mathrm{C,1}$. As $A_\mathrm{BAW}$ is increased, the narrow resonances broaden symmetrically. As is expected for the strain modulation, the maximum value of the QD energy excursion $\Delta E (A_{BAW})$ (indicated by dashed lines) with respect to the energy without modulation is a linear function of $A_\mathrm{BAW}$. Essentially, the modulated QD resonances map out photonic modes. 

We observe that the $QD_2$ energy crosses $E_\mathrm{C,1}$ for $A_\mathrm{BAW} \approx 0.25$, while the $QD_3$ crosses $E_\mathrm{C,0}$ for $A_\mathrm{BAW} \approx 0.35$. It indicates that the $QD_3$ resonance is slightly less sensitive to the BAW strain. This could indicate a reduced spatial overlap between the QD position and the phonon mode of the trap. For the largest measured $A_\mathrm{BAW} = 0.56$, the maximum $\Delta E$ reaches 3~meV. Intensity profiles for $QD_2$ and $QD_3$, taken along the dashed lines, are shown in Fig.~\ref{Fig3}c. 
%
    \hBlue{From intensity profiles of Fig.~\ref{Fig3}c, we see that the intensity of a QD is reduced when it is shifted between the optical modes and increases on resonance. Phenomenologically same behavior was observed by Weiß et al.~\cite{weiss2016Appl.Phys.Lett.}, where the QD resonance was modulated by an 800 MHz surface acoustic wave, as well as by Pagliano et al.~\cite{pagliano2014NatCommun} for a Stark-modulated QD in a cavity.}
These observations constitute the first demonstration of a dynamic Purcell suppression and enhancement at triggering frequencies exceeding a GHz.

    \hBlue{ 
    We can assess the effect of the time-periodic dynamical crossing on the QD exciton population and photon emission by evaluating the decay of the exciton under the time-dependent spontaneous emission rate due to the dynamical Purcell effect. Previously, such analysis has been carried out by Thyrrestrup et al.~\cite{thyrrestrup2013Opt.Express} for a case of a fast switching of the QD-cavity resonance followed by a slow energy-decay of the cavity mode. In our case, we have a time-periodic continuous evolution of the Purcell factor due to the strain modulation of QD-cavity detuning: $ { \Delta E_{XC}(t) = E_{C} - E_{X}(t) }$, where the time-independent cavity energy is $ { E_{C} }$ and time-dependent exciton energy is $ { E_{X}(t) = E_{X,0} - A_{Mod} Cos(2 \pi F_{BAW} t) }$, with $ { E_{X,0} }$, $ { A_{Mod} }$ and $ { F_{BAW} }$ being static exciton energy, energy modulation amplitude and BAW modulation frequency, respectively. In principle, by tuning the modulation amplitude we can study two scenarios: (i) the modulation amplitude ($ { A_{Mod} }$) is equal to the static QD-cavity detuning, i.e., $ { A_{Mod} = \Delta E_{XC,0} }$, where $ { \Delta E_{XC,0} = E_{C} - E_{X,0} }$, leading to the fast pivot at resonance and (ii) a rapid passage, when $ { A_{Mod} > \Delta E_{XC,0} }$. The rapid passage case, is similar to the one considered in Ref.~\cite{peinke2021LightSci.Appl.}. In the following, we will focus on the case (i) $ { A_{Mod} = \Delta E_{XC,0} }$. This case presents very clear differences to the conclusions in Ref.~\cite{peinke2021LightSci.Appl.}.
   
    Figure~\ref{Fig4}a shows a generalized time-evolution of the exciton energy for $ { A_{Mod} = \Delta E_{XC,0} }$ over three modulation periods. To evaluate exciton and emission dynamics, we need to calculate the time-evolution of detuning-dependent Purcell factor, $ { F_{P}(t) }$. According to Ref.~\cite{englund2005Phys.Rev.Lett.}, its value is given by $F_{P}(t) = \frac{ 3 Q_{C} }{4 \pi^2 V_{C}} * (\frac{\lambda_{C}}{n_{C}})^3 * \frac{1}{1 + 4[(E_{X}(t) - E_{C})/\gamma_{C}]^2}$, where $ V_{C}$ is the optical mode volume and $ {\gamma_{C}}$ is the cavity linewidth. As shown in the work~\cite{sultanivala2025J.Phys.Photonics}, optical volumes as small as $ {V_{C} \approx 20 \times (\lambda_{C}/n_{C})^3}$ are possible in etch-and-overgrowth traps. Based on the approach in Ref.~\cite{koks2021Opt.Express}, we estimated the mode volume for our MC and $ {1 \times 1~\mu m^2}$ trap to be $ {V_{C} \approx 40 \times (\lambda_{C}/n_{C})^3}$. The procedure is described in SI-Sec.~10.

    \begin{figure}[!t]
    	\centering
    		\includegraphics[width=0.5\textwidth, keepaspectratio=true]{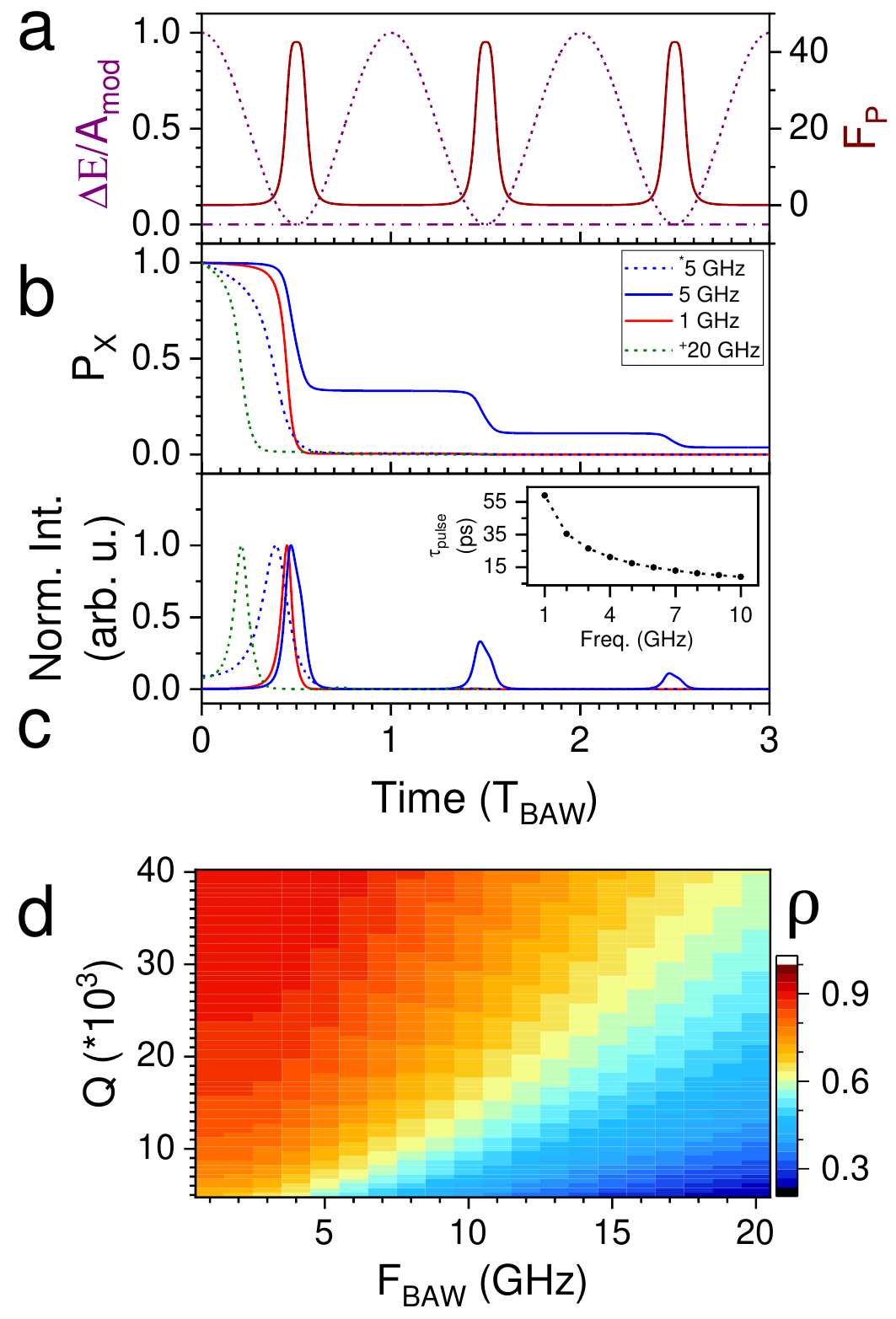}
    		\caption{
                {\bf Simulated time-evolution under modulation.}
                {\bf a} Energy vs. time for the exciton (dashed line) and photon (dash-dotted line) and Purcell factor (solid line). The energy scale is relative to the photon energy and normalized to the modulation amplitude ($ { A_{Mod} }$).  
                {\bf b \& c} Evolution of the QD exciton population b and intensity c for different frequencies and conditions as described in text. The time scale is in units of the modulation period $ { T_{BAW} }$.
                {\bf d} Map of the figure of merit $ { \rho }$ (defined in the text) as a function of Q and $F_{BAW}$. 
               }
    	\label{Fig4}
    \end{figure}

    First, we consider the case of $ { F_{BAW} = 5 ~GHz}$ and  $ { A_{Mod} = \Delta E_{XC,0} = 0.5 ~ meV }$, which is comparable to the experimental values of Fig. 3a over a broad range of frequencies. We assume a $ {Q_{C} = 20000}$ (corresponding to the simulated Q of the cavity, SI-Sec. 1) giving cavity linewidth of $ {\gamma_{C} = 0.065 ~ meV}$. 

    Figure~\ref{Fig4}a also shows simulated time-evolution of the Purcell factor for a $ {1 \times 1~\mu m^2}$ trap.  As expected, when the exciton is approx. within the cavity linewidth, there is a rapid increase of the Purcell-factor value, which repeats with the modulation period $ {T_{BAW} = 1/F_{BAW}}$. The maximum Purcell enhancement reaches $ {F_{P} \approx 45}$. We note that for the MC structure and mesas with parameters similar to our work, e.g., diameters of $ {1 ~ \mu m}$ to $ {3 ~ \mu m}$, mesa height of 13 nm, number of top-DBR optical pairs $ {N_{tDBR} = 33}$, $ {F_{P}}$ up to 70 have been predicted~\cite{sultanivala2025J.Phys.Photonics}.

    We then calculate time-dependent exciton population $ { P_{X}(t) = P_{X,0} * Exp[-\Gamma(t)] }$, where $ { P_{X,0} =1 }$ is the exciton population at zero time and the time-integrated decay rate into the mode is $ { \Gamma(t) = 1/\tau_{bare} \int_{0}^{t} F_{P}(t^*) \,dt^* }$, where $ {\tau_{bare}}$ is the exciton isotropic radiative lifetime. Considering only the radiative channel, the intensity evolution is then given by $ {I(t) = \Gamma_{rad}(t) P_{X}(t)}$, where the instantaneous radiative decay rate is given by $ { \Gamma_{rad}(t) = F_{P} / \tau_{bare} }$. In our analysis, we assumed that collection efficiency is unit~\cite{peinke2021LightSci.Appl.}. It has been shown that the $Q$-factor of the etch-and-overgrowth trap remains high even below $1~\mu$m, whereas $Q$ of etched pillar cavities deteriorates considerably~\cite{karl2009Opt.Express, sultanivala2025J.Phys.Photonics}. Therefore, we also neglected possible contribution due to the exciton spontaneous emission into non-resonant continuum of states. 

    The solid red and blue curves in Fig.~\ref{Fig4}(b) show $ { P_{X}(t) }$ for $ { F_{BAW} = 1~GHz }$ and $ { F_{BAW} = 5~GHz }$, respectively, assuming $ {\tau_{bare} = 1000 ~ ps}$ and $ { A_{Mod} = \Delta E_{XC,0} = 0.5 ~ meV }$. For $ { F_{BAW} = 1~GHz }$, exciton population rapidly reduces to zero during the first dynamic QD-cavity resonance at $ { t = 0.5 T_{BAW} }$. This corresponds to the emission of a single short pulse shown by the solid red curve in Fig.~\ref{Fig4}c. The SP pulse has a Lorentzian shape with the pulse width given by the full-width at half-maximum of $ { \tau_{SP} = 59 ~ ps }$.  For $ { F_{BAW} = 5~GHz }$, we find, after the first resonance, $ { P_{X}(0.7 T_{BAW}) \approx 0.33}$, followed by further sharp drops at $ { t = 1.5 T_{BAW} }$ and ${ t = 2.5 T_{BAW} }$, at which point $ { P_{X}(2.7 T_{BAW}) \approx 0.03}$. The $I(t)$ for this case, shown by the solid blue curve in Fig.~\ref{Fig4}c, reveals three short pulses, with the fist and the strongest pulse having ${ \tau_{SP} = 17.5 ~ ps }$. This demonstrates that in both cases, the sub 100 ps single-photon pulses will be emitted. The inset of Fig.\ref{Fig4}c shows a dependence of ${ \tau_{SP}}$ on frequency for the above parameters. We note that at $F_{BAW} = 5 ~ GHz$ there is a non-zero emission up to two consecutive resonances, which could reduce SP purity of the source. Strategies to reach complete radiative decay within one dynamical resonance for frequencies exceeding 5 GHz will be discussed in the following section of the paper. 
   
    We point out a difference between $ { A_{Mod} = \Delta E_{XC,0} }$ and $ { A_{Mod} > \Delta E_{XC,0} }$ cases. For the case $ { A_{Mod} = \Delta E_{XC,0} }$, the maximum relative change of the exciton population $\Delta P_{X}$ after a single pivot at QD-cavity resonance depends on the cavity Q-factor. For the case $ { A_{Mod} > \Delta E_{XC,0} }$, we obtain behavior identical to the one in Ref.~\cite{peinke2021LightSci.Appl.} – the single-passage $\Delta P_{X}$ is independent of Q-factor. Relevant simulations are shown in SI-Sec.~11.
    }

\section{Discussion}
\label{Discussions}

In summary, we have integrated InAs QDs into the spacer of a hybrid photon-phonon patterned AlGaAs MC with $\mu$m-sized photonic traps. Small traps were found to contain as few as a single QD in a sample region with low QD density. The SP emission from the QD in a trap was confirmed by measuring the value of $g^{(2)}(0) < 0.5$. We have shown that the dynamical strain of a GHz BAW, piezoelectrically injected into a trap using a BAR, couples to the QD transition energy via the DP mechanism leading to the modification of the QD PL spectrum. Due to the large bandwidth of the BAR, QD spectrum has been harmonically modulated with a frequency up to 14~GHz and considerable energy modulation amplitudes reaching 3~meV and exceeding the energy spacing between confined optical modes. At $F_\mathrm{BAW} = 5.6$~GHz, the dynamic crossing between the QD energy and a spectrally narrow confined cavity mode leads to an appreciable enhancement of PL, whereas PL is suppressed when off-resonance, demonstrating the dynamic Purcell suppression and enhancement of SP emission triggered at 5.6~GHz. 
    \hBlue{We numerically studied time-evolution of the QD-cavity system under modulation and identified conditions that could enable single-photon rates exceeding 5~GHz.} 
Our work has demonstrated a promising platform for scalable and compact on-demand deterministic SPSs with triggered SP rates in the GHz-range under continuous wave optical excitation, where the GHz BAWs act as an on-chip GHz clock for the SP emission.

    \hBlue{Precise engineering of both optical and acoustic degrees of freedom enabled us to realize modulation of QD energy up to 14 GHz frequency and dynamical Purcell effect at around 6 GHz, demonstrating considerable improvement over previous realizations. Table~\ref{table_Comparison} compares the maximum reported modulation frequency ($F_{max.}$) and the optimum frequency ($F_{optm.}$, for which the dynamical Purcell effect has been robustly demonstrated) for different approaches. 
    For example, in contrast to Ref.~\cite{pagliano2014NatCommun}, where the dynamical characteristics of the devices showed reduction of the modulation amplitude and deterioration of the temporal waveform at high frequencies, our approach can be extended to 20~GHz by straightforward modification of the cavity structure~\cite{kuznetsov2021Phys.Rev.X}.
    \hBlue{Interestingly, as can be seen in Fig. 3b, for ${ A_{BAW} > 0.35 }$, the ${QD_2}$ energy shifts above ${ E_{C,1} }$, which means that it passes ${ E_{C,1} }$ two times per cycle effectively doubling the triggering rate.}
    
    \begin{table}[h]
      \centering
    \begin{threeparttable}
      \caption{\hBlue{Comparison of realizations of dynamical Purcell effect.}}
      \begin{tabular}{| c | c  | c | c |}
     \hline
     \hline
     Modulation type  & $F_{max.}$ (GHz) & $F_{optm.}$ (GHz)   &   Ref. \\
     \hline
     Optical pulse\tnote     &   0.08     &   0.08    & \cite{peinke2021LightSci.Appl.}  \\ 
     \hline
     Acoustic wave\tnote{a}     &   0.8      &    0.8   & \cite{weiss2016Appl.Phys.Lett.}  \\ 
     \hline
     Electric field\tnote     &   3     &   0.3    & \cite{pagliano2014NatCommun}  \\ 
     \hline
     Acoustic wave\tnote{b}     &   14     &   5.6\tnote{c}    & This work  \\ 

    \hline    
    \hline
    \end{tabular}
      \begin{tablenotes}
        \item[a] Surface acoustic wave. 
        \item[b] Bulk acoustic wave. 
        \item[c] Acoustic cavity mode.   
      \end{tablenotes}
    \label{table_Comparison}
    \end{threeparttable}
    \end{table}
    } 

    \hBlue{In order to accurately quantify dynamical enhancement and suppression, determination of the static Purcell enhancement via lifetime measurements will need to be performed on high-Q samples. The decay rate constant of 1 ns obtained from fitting the $g^{(2)}$ data of Fig.~\ref{Fig2}d, provides a lower limit on the radiative lifetime of a QD in a trap, which is very close to the measured radiative lifetime of 1~ns of similar QDs in Ref.~\cite{decrescent2024Optica}. Hence, in the present case, Purcell enhancement factor is $F_{P} \approx 1$, which can be expected for the reduced value of $Q \approx 1500$ of our pilot sample.}
    \hBlue{Therefore, in the present case (low-Q sample), the demonstrated modulation will lead to an amplitude modulation instead of on/off emission of SPs. As a mitigation strategy, an external narrow-band optical filter can be introduced to improve the modulation depth~\cite{peinke2021LightSci.Appl.}.}
We note that similar MCs with quantum wells typically have $Q \approx 7500$, while in the present case $Q \approx 1500 \pm 100$. This discrepancy is most likely due to the MBE overgrowth parameters (e.g., substrate temperature and surface cleaning before the overgrowth) being not optimized for the new type of sample. Beyond improving the Q factor for the generation of SP pulses, the next generation of the MC would include doped layers for the charge stabilization~\cite{wang2024Opt.Express}, which would enable $g^{(2)}(\tau)$ measurements under resonant optical excitation. 
    \hBlue{The fluctuating charge environment around a QD is responsible for spectral diffusion. Well-defined charge environment prevents random hopping between different charge states~\cite{martin2025Appl.Phys.Lett.} leading to better multiphoton suppression, which has a direct impact on single-photon statistics~\cite{chen2016Phys.Rev.B}.}

    \hBlue{We highlight an important advantage of acoustic control. As has been demonstrated by Weiss et al.~\cite{weiss2016Appl.Phys.Lett.}, the dynamic Purcell effect induced by acoustic modulation maintains SP character of emission. Similarly, other reports show that GHz acoustic modulation of two-level systems does not introduce decoherence~\cite{Zhan2026Phys.Rev.Lett.a,groll2026J.Phys.Photonics}.}



    \hBlue{In the following, we discuss complementary strategies to improve SP purity at higher frequencies. 
    The first strategy is to work with small static QD-cavity detuning in the condition $ { A_{Mod} = \Delta E_{XC,0} }$, which would lead to longer QD-cavity resonance, and to select QDs with shorter $\tau_{bare}$. For example, for  $\tau_{bare} = 500 ~ ps$,  $ { A_{Mod} = \Delta E_{XC,0} = 0.1 ~ meV }$ and $Q = 20000$, we predict full switching at the first dynamical resonance at 5 GHz, as shown in the Fig.~\ref{Fig4}b by the dashed blue line. We note that $\tau_{bare} = 670 ~ ps$ have been measured for InAs QDs~\cite{mannel2026J.Appl.Phys.}. Much higher frequencies can be achieved using GaAs QDs, where exciton $100~ps \leq \tau_{bare} \leq 200~ps$ have been demonstrated~\cite{hours2005Phys.Rev.B}. For the latter case, assuming $\tau_{bare} = 150 ~ ps$ and $ { A_{Mod} = \Delta E_{XC,0} = 0.1 ~ meV }$, we predict single pulse emission up to $F_{BAW} = 20 ~ GHz$, as shown by the dashed green curve in Fig.~\ref{Fig4}b. We have demonstrated BAW modulation of exciton resonances in QWs in similar MCs at 20 GHz~\cite{kuznetsov2021Phys.Rev.X}. We point out that in the case of small $A_{Mod}$, the trade-off is the increased SP pulse duration relative to $T_{BAW}$, as shown in Fig.~\ref{Fig4}c by dashed curves for the 5 GHz and 20 GHz cases. 

    We note that in all above cases, the Q-factor was constant $Q = 20000$. Now, we consider the second strategy to increase frequency to generate single photon pulses, which is related to increasing the Q-factor of the cavity. We set $\tau_{bare} = 500 ~ ps$, $ { A_{Mod} = \Delta E_{XC,0} = 0.25 ~ meV }$ and calculate a figure of merit, which quantifies both the radiatively decayed exciton population after the first dynamical resonances and the intensity pulse duration, i.e., $\rho = (1 - P_{X})(1 - D)$, where $D = \tau_{SP} / T_{BAW}$ is the duty cycle. Hence, high values of $\rho$ characterize simultaneously high conversion efficiency and short SP pulses. Figure~\ref{Fig4}d shows a 2D map of $\rho$ as a function of $F_{BAW}$ and Q. We see that below 5 GHz, the high values of $\rho$ are achieved for $Q \approx 10000$.

    The third strategy is to use one photonic mode, e.g., $E_{C,i+1}$, for CW excitation and the other mode (e.g., $E_{C,i} < E_{C,i+1}$) for SP emission, while adjusting the $A_{Mod} \geq E_{C,i+1} - E_{C,i}$, cf. sketch of Fig.~\ref{Fig1}a. In this case, both the excitation of the QD and SP emission will be precisely clocked at the $F_{BAW}$ and separated in time, which would get rid of the temporal overlap of consecutive SP pulses. Another advantage of this strategy is that the quasi-resonant excitation and emission will be separated spectrally, thus eliminating scattered laser photons from the SP signal.

    We note that the above strategies are complementary, which, we believe, could considerably lower the requirements for reaching very high triggered SP rates in comparison to each individual approach.}
    \hBlue{In order to reach SP rates potentially exceeding 10 GHz, we could use GaAs QDs, which have very short isotropic radiative lifetime below 200 ps~\cite{hours2005Phys.Rev.B}, therefore enabling triggered SP emission rates in the GHz range even for modest Purcell enhancements of $F_{P} \approx 10$.}

    \hBlue{In contrast to complex and costly external EOM modulation or pulse multiplication techniques, our approach is fully on-chip and enables straightforward scaling to higher frequencies by reducing piezoelectric film thickness or using high acoustic velocity piezoelectric materials, e.g., Al(Sc)N. Piezoelectric transducers generating at frequencies exceeding 50~GHz have recently been demonstrated~\cite{kramer2025Appl.Phys.Lett.}. 
    Furthermore, since the footprint of acoustic transducers is small (sub-mm), and in combination with precise~\cite{madigawa2024ACSPhotonics} deterministic fabrication of photonic traps with single QDs~\cite{rodt2021NanoExpress}, many devices can be fabricated on the same chip and operated independently or synchronously. The latter regime offers a unique possibility of synchronization of remote SP sources to produce streams of single-photons with tunable temporal configurations.    
    One future development would be an integration of a CW laser on the same chip for ultimate integration and miniaturization.}



\section*{Methods}

\textbf{QD microcavity sample}

The basic design of the MC has been detailed in Refs.~\cite{kuznetsov2023Nat.Commun.,kuznetsov202a6Nat.Photon.}. In short, here we rely on the ``double magic coincidence''~\cite{trigo2002Phys.Rev.Lett.,fainstein2013Phys.Rev.Lett.} exhibited in (Al,Ga)As systems between light and GHz acoustic vibrations. The latter means that the spacer of the MC co-localizes optical fields of approximately 300~THz and longitudinal acoustic strain in the GHz range. The planar MC is expected to have Q = 5000. The spacer of the MC contains a single layer of self-assembled InAs QDs emitting single photons in the 900--950~nm range. The whole MC structure is designed such that QDs are located very close to both photon and phonon anti-nodes. The strain predominantly couples to QDs via the DP mechanism~\cite{kuznetsov2021Phys.Rev.X}.

First, the molecular beam epitaxy (MBE) is used to grow the lower DBR and the cavity spacer with QDs. The spacer is then subjected to shallow etching (etching depth of approx. 10~nm) with micrometer lateral scale. Etching takes place at least 200~nm away from the QD depth. The sample is then re-inserted into the MBE chamber and the upper DBR is overgrown. The patterning leads to the lateral confinement of the photonic mode. No post-selection of QDs is done.

\textbf{Bulk acoustic wave transducers}

The acoustic mode at $F_\mathrm{M} \approx 5.6$~GHz of the MC was resonantly driven by a BAR~\cite{machado2019Phys.Rev.Appl.} fabricated on top of the region containing the trap and excited using an RF generator, cf. Fig.~\ref{Fig1}a. The longitudinal BAW is excited via the piezoelectric effect by the BAR away from the trap. The BAW reaches the center of the trap after lateral propagation involving multiple reflections between the polished bottom and top sides of the sample. The amplitude of the BAW can be precisely controlled by the RF power applied to BAR. A signal-ground RF probe was used to connect the bottom and top electrodes of the transducer to an RF generator with high-power output (up to 25~dBm).

\textbf{Optical measurements under modulation}

Optical measurements under acoustic excitation were carried out in a cold-finger liquid-He cryogenic probe station at 5~K temperature. BARs were electrically contacted using RF probes. A single-mode continuous wave external cavity-stabilized Ti-Sapphire laser was used to non-resonantly excite QDs for PL measurements with BAW modulation. The laser was focused on the sample at normal incidence using a 10x objective resulting in a few-$\mu$m Gaussian-like spot positioned on the trap. PL measurements integrated over 0.1--1 second with moderate spectral resolution of 0.1~meV, were carried out by transferring PL image of the trap on the entrance slit of a single grating spectrometer and recorded using a nitrogen-cooled CCD camera.

The second-order autocorrelation measurements $g^{(2)}(\tau)$ are performed using a setup similar to that described in Ref.~\cite{decrescent2024Optica}. The sample is mounted in a closed-cycle cryostat and cooled to approximately 5 K. Excitation is provided by a 640 nm laser focused onto the sample through a microscope objective with a numerical aperture of 0.81. The reflected laser light is separated from the PL signal using an 850 nm long-pass filter. A single emission line is then spectrally isolated with a narrow band-pass filter of about 2 nm bandwidth. The filtered PL is directed to a 50:50 non-polarizing beam splitter, and the two output ports are coupled to superconducting nanowire single-photon detectors. Photon arrival times from the two detectors are recorded by a picosecond-resolution event timer \hBlue{(time resolution 64 ps)}, and the coincidence histogram is constructed using standard time-correlated single-photon counting techniques to obtain $g^{(2)}(\tau)$.

\section*{Data availability}
The measurement and numerical simulation data that support the findings within this study are included within the main text and Supplementary Information and can also be made available upon request from the corresponding author.

\section*{Acknowledgments} 
A.S.K. acknowledges the funding from DFG grant 359162958. 
The authors thank Dr. Valentino Pistore for a critical review of the manuscript. 

~\\
\section*{Author contributions}

A.S.K. performed the spectroscopic experiments with acoustic modulation and carried out the respective analysis. 
M.S. fabricated bulk acoustic transducers.
Z.W. carried out the photon correlation measurements. Sample design was done by K.B. with the input of all authors. 
K.L.S. fabricated the lower half of the MC with QDs. 
K.B. performed the MBE overgrowth of the samples. 
A.S.K. has conceived the idea and written the manuscript with the input from all authors.

\section*{Competing interests}
The authors declare no competing interests.


\bibliography{ZoteroBib}



\end{document}


\title{Supplementary Information:\\
Deterministic single-photon source with on-chip 5.6 GHz acoustic clock}

\author{Alexander S. Kuznetsov}
\email[corresponding author: ]{kuznetsov@pdi-berlin.de}
\affiliation{Paul-Drude-Institut f{\"u}r Festk{\"o}rperelektronik, Leibniz-Institut im Forschungsverbund Berlin e.V., Hausvogteiplatz 5-7, 10117 Berlin, Germany}

\author{Meysam Saeedi}
\affiliation{Paul-Drude-Institut f{\"u}r Festk{\"o}rperelektronik, Leibniz-Institut im Forschungsverbund Berlin e.V., Hausvogteiplatz 5-7, 10117 Berlin, Germany}

\author{Zixuan Wang}
\affiliation{Department of Physics, University of Colorado, Boulder, Colorado 80309, USA}
\affiliation{National Institute of Standards and Technology, Boulder, Colorado 80305, USA}

\author{Kevin L. Silverman}
\affiliation{National Institute of Standards and Technology, Boulder, Colorado 80305, USA}

\author{Klaus Biermann}
\affiliation{Paul-Drude-Institut f{\"u}r Festk{\"o}rperelektronik, Leibniz-Institut im Forschungsverbund Berlin e.V., Hausvogteiplatz 5-7, 10117 Berlin, Germany}

\maketitle

\vskip 1 cm
\begin{center}
This document contains additional material to support the conclusions of the main text.
\end{center}
\vskip 1 cm

\tableofcontents



\newpage
\section{Microcavity optical and acoustic properties}
\label{appendix:id}

\justifying

The microcavity (MC) design is based on the one in Ref.~\cite{kuznetsov2023Nat.Commun.}, where we studied acoustic modulation of light-matter exciton-polaritons. We take advantage of the optical and acoustic coincidence of AlGaAs~\cite{fainstein2013Phys.Rev.Lett.}, where an AlGaAs MC designed for photons in the 800--900 nm range acts as a MC for GHz longitudinal bulk acoustic phonons. The MC consists of a spacer, which contains a delta-like layer of InAs self-assembled quantum dots (QDs), sandwitched between a lower (LDBR) and an upper (UDBR) distributed Bragg reflectors (DBRs). Each DBR period consists of a stack of three pairs of  Al$_\mathrm{x_1}$Ga$_\mathrm{(1-x_1)}$As/Al$_\mathrm{x_2}$Ga$_\mathrm{(1-x_2)}$As layers, each with an optical thickness $\lambda/4$. The thicknesses $d_{1}$ and $d_{2}$ and Al compositions  $x_{1}$ and $x_{2}$ of the layers pairs are listed in Table \ref{SMTable1}.

\begin{table}[htbp]
  \caption{Layer structure of the MC sample. The DBRs consist of stacks of Al$_{x_1}$Ga$_{1-x_1}$As and Al$_{x_2}$Ga$_{1-x_2}$As layers with thicknesses  $d_1$ and $d_2$ and Al compositions $x_1$ and $x_2$, respectively.  $d_T$ and $n_\mathrm{rep}$ are, respectively, the total thickness and the number of repetition periods.}
  \begin{tabular}{|| c | c  | c | c  | c   | c  || }
 \hline
 \hline
 Region   &$d_T$ (nm)& $n_\mathrm{rep}$   &   $d_1:d_2$ (nm) & $x_1:x_2$ & Comment \\
 \hline
         & 4978.6   & 11        & 76.9:69.5    & 0.6:0.05 & pair$_3$ \\   
 UDBR     &         &           & 81.2:69.5    & 0.95:0.05 & pair$_2$ \\
          &         &           & 81.2:74.3    & 0.95:0.4 & pair$_1$ \\
 \hline
       & 69.5     & 1         &   69.5          & 0.05      &  \\
         &  ---     &  ---           & ---       &  ---   &  growth interruption  \\
        & 197.9     & 1         &   197.9          & 0.05      & \\
 spacer & 10     & 1         &   10          & 0      & \\
        &   ***   &      ***    &      ***       &   ***    & layer of InAs QDs \\
        & 10     & 1         &   10          & 0      & \\
        & 129.3     & 1         &   129.3          & 0.05      & \\
 \hline
       & 6336.4    & 14        & 76.9:69.5    & 0.6:0.05 & pair$_3$ \\   
 LDBR   &         &           & 81.2:69.5    & 0.95:0.05 & pair$_2$ \\
          &         &           & 81.2:74.3    & 0.95:0.4 & pair$_1$ \\
  \hline
 substrate&         &           &                &           &  GaAs(001) S-I substrate (double side polished) \\
\hline    
\hline
\end{tabular}
\label{SMTable1}
\end{table}

\begin{figure*}[h]
	\centering
	\includegraphics[width=0.9\textwidth, keepaspectratio=true]{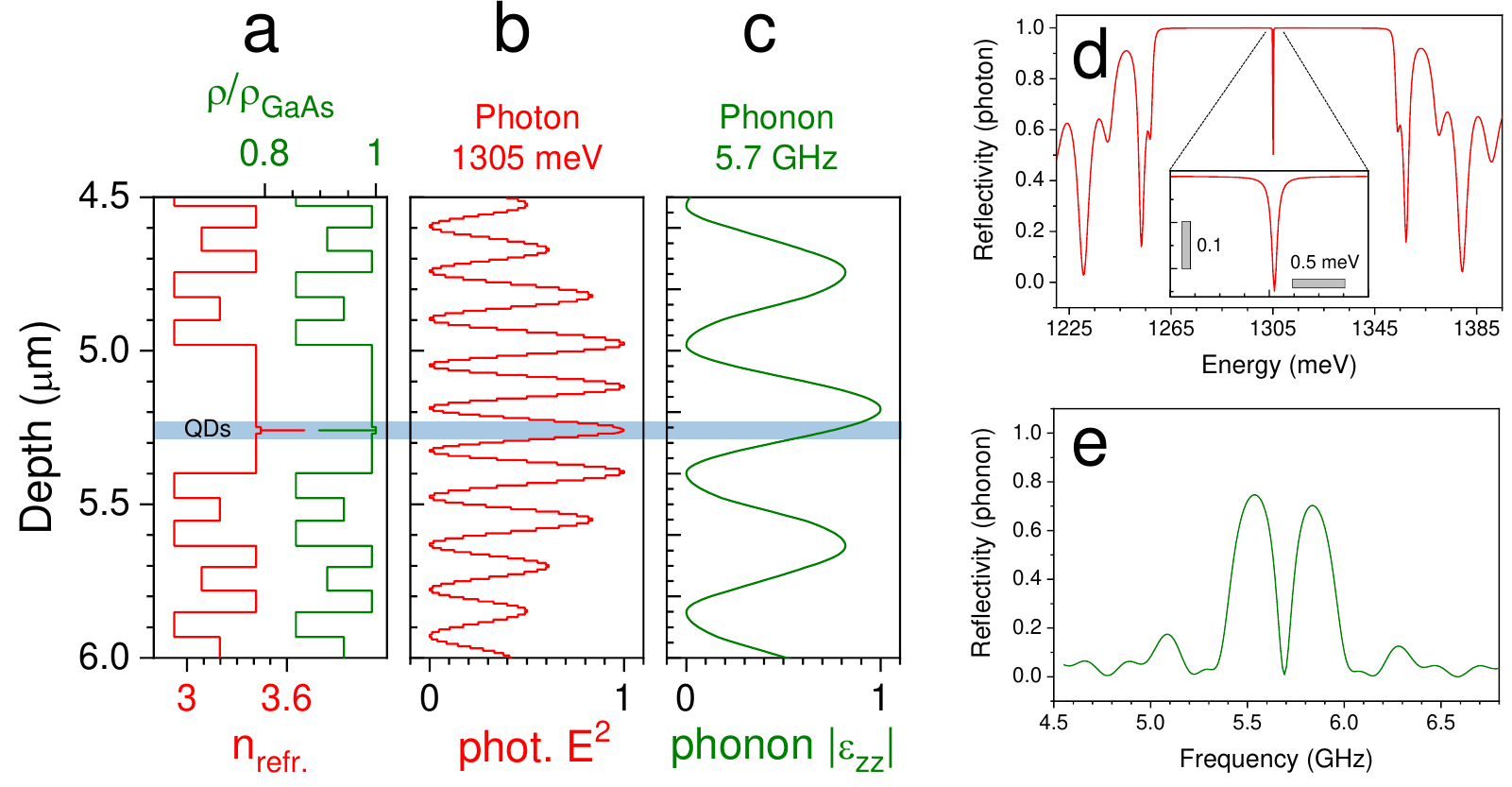}
	\caption{
        {\bf Simulated microcavity optical and acoustic responses.}
        {\bf a} Depth profiles of the refractive index and normalized mass density, {\bf b} optical field at 1530 meV and {\bf c} amplitude of 5.6~GHz phonon strain around the MC spacer (zero at the surface).
        {\bf d,e} Calculated optical and acoustic reflectivity spectra, respectively. The inset in {\bf d} shows a zoom of the optical cavity mode.  
        }
\label{Fig_SI_MC-Simulations}
\end{figure*}

Figure~SI~\ref{Fig_SI_MC-Simulations}a shows the depth profiles of the refractive index and the normalized acoustic impedance of a region around the MC spacer. These curves show that the layers will act in an similar matter on both a $\sim 950$~nm photon and a few GHz acoustic phonon modes. Curves of Figs.~SI~\ref{Fig_SI_MC-Simulations}b,c depict the depth profiles of the optical and acoustic fields calculated using a transfer matrix approach. The spacer layer acts a $3/2 \lambda_\mathrm{photon}$ optical and $\lambda_\mathrm{phonon}$ acoustic cavity. When designing the MC, we make sure that the QD layer is positioned close to both optical and acoustic field anti-nodes. This design maximizes the deformation potential coupling of the acoustic strain to the QD excitons, which dominates over any photoelastic coupling~\cite{sesin2023Phys.Rev.Res.}. 

Figures~SI~\ref{Fig_SI_MC-Simulations}d,e show simulated optical and acoustic reflectivities, respectively. Both plots exhibit clear stopband regions with well-defined cavity modes at 950 nm (photons) and 5.7 GHz (phonons). Due to the lower number of acoustic DBR pairs, the acoustic cavity resonance is broad, which translates in an acoustic quality factor of $Q_\mathrm{a}\approx 70$, while the optical one is $Q_\mathrm{o}\approx 20000$, as can be seen in the inset of Fig.~SI~~\ref{Fig_SI_MC-Simulations}d. 

\newpage
\section{Spatial reflectivity}
\label{appendix:reference}

\begin{figure*}[h]
	\centering
	\includegraphics[width=0.9\textwidth, keepaspectratio=true]{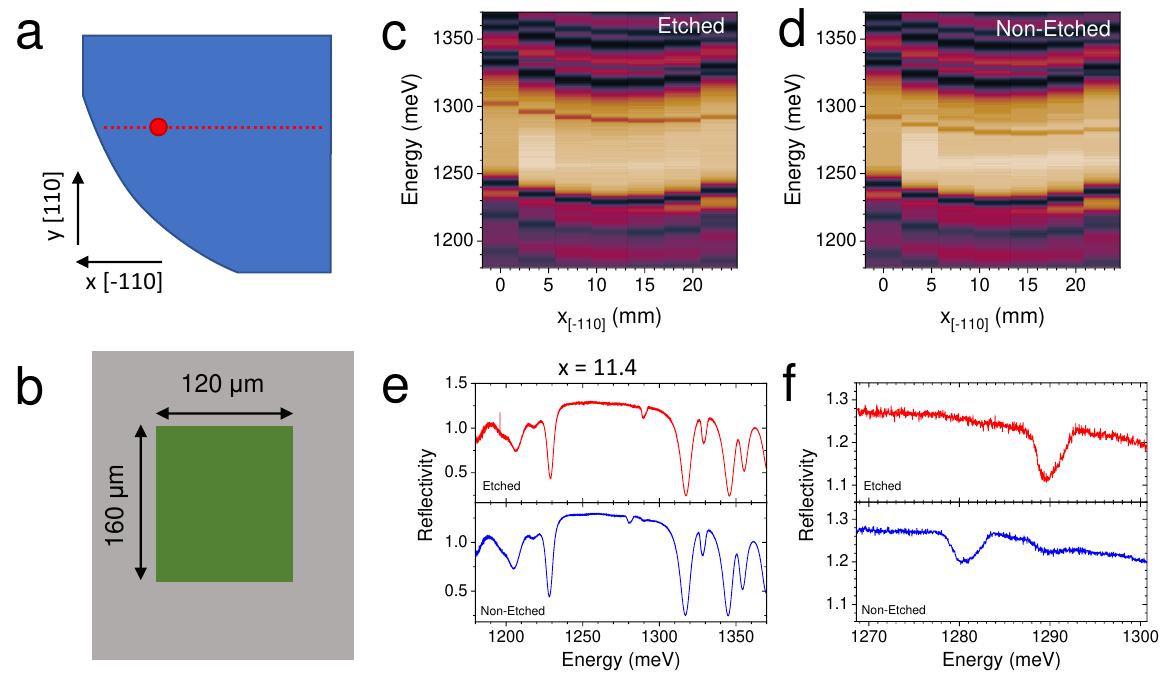}
	\caption{
        {\bf Spatial optical reflectivity.}
        {\bf a} A sketch of the microcavity sample. The red dot corresponds to the $x = 0$~mm position.
        {\bf b} A sketch of the area used to measure the reflectivity of the non-etched (gree rectangle) and etched (gray shaded area) regions.
        {\bf c,d} Reflectivity spectra of spatially extended etched and non-etched MC regions measured across the sample at 5~K. The zero value corresponds to the red point in $a$. The bright colors correspond to higher reflectivity.
        {\bf e} Single spectra of the etched and non-etched regions at x = 11~mm. The values of the reflectivity larger than unit are an artifact of normalization to the reflectivity reference. 
        {\bf f} Zoom of the spectra in $c$ in the range of the cavity modes.
        }
\label{Fig_SI_SpatialReflectivity}
\end{figure*}

Here we discuss spatial dependence of MC reflectivity. The studied microcavity was grown on a quarter of a 3" GaAs wafer as schematically shown in Figures~SI~\ref{Fig_SI_SpatialReflectivity}a. There, the dashed red line shows the spatial direction of the reflectivity scan. Reflectivity of the non-etched and etched regions have been measured on a structure that is schematically shown in  Figures~SI~\ref{Fig_SI_SpatialReflectivity}b. It consists of a $160 \times 120 \mu \mathrm{m^2}$ non-etched (nER) mesa surrounded by etched (ER) area. The large areas of the nER and ER ensure that no lateral confinement of photonic modes is present. 

Figures~SI~\ref{Fig_SI_SpatialReflectivity}c,d show spatial reflectivity of non-etched and etched areas, respectively measured at 5~K. Both maps show very clear parabolic spatial dispersion of the optical resonances, characteristic of the MBE growth under rotation.  

Figure~SI~\ref{Fig_SI_SpatialReflectivity}e depicts exemplary nER and ER curves for a fixed measurement position. In both curves, a flat region in the reflectivity corresponds to an optical stopband. In the stopband, there is a single narrow resonance due to the optical mode. A zoom-in of the optical mode, cf. Figure~SI~\ref{Fig_SI_SpatialReflectivity}f, shows that the ER resonance is blueshifted by approximately 10~meV with respect to the one in the nER, which is the consequence of the etching of the MC spacer. The magnitude of the energy shift directly translates to the depth of the confinement potential as will be discussed in the next section.

We note that the linewidth of the optical resonances is artificially broadened due to the numerical aperture of the 10x objective used in the measurement. The weak signature of the ER optical mode in the ER spectrum is due to the fact that the reflectivity sampling area is larger than the size of the nER region.

\newpage
\section{Photonic confinement}
\label{appendix:reference}

The lateral zero-dimensional photonic confinement is introduced by creating micron-scale lateral mesas in the spacer of the microcavity by shallow (approx. 13 nm) etching followed by the MBE overgrowth~\cite{kuznetsov2018Phys.Rev.B}. As was discussed in the previous section, such etching introduces up to 10 meV energy shifts between the cavity resonances in the etched and non-etched regions. The size and the shape of the mesa, which acts as a trap, then dictates the structure of the confined levels.

\begin{figure*}[h]
	\centering
	\includegraphics[width=0.9\textwidth, keepaspectratio=true]{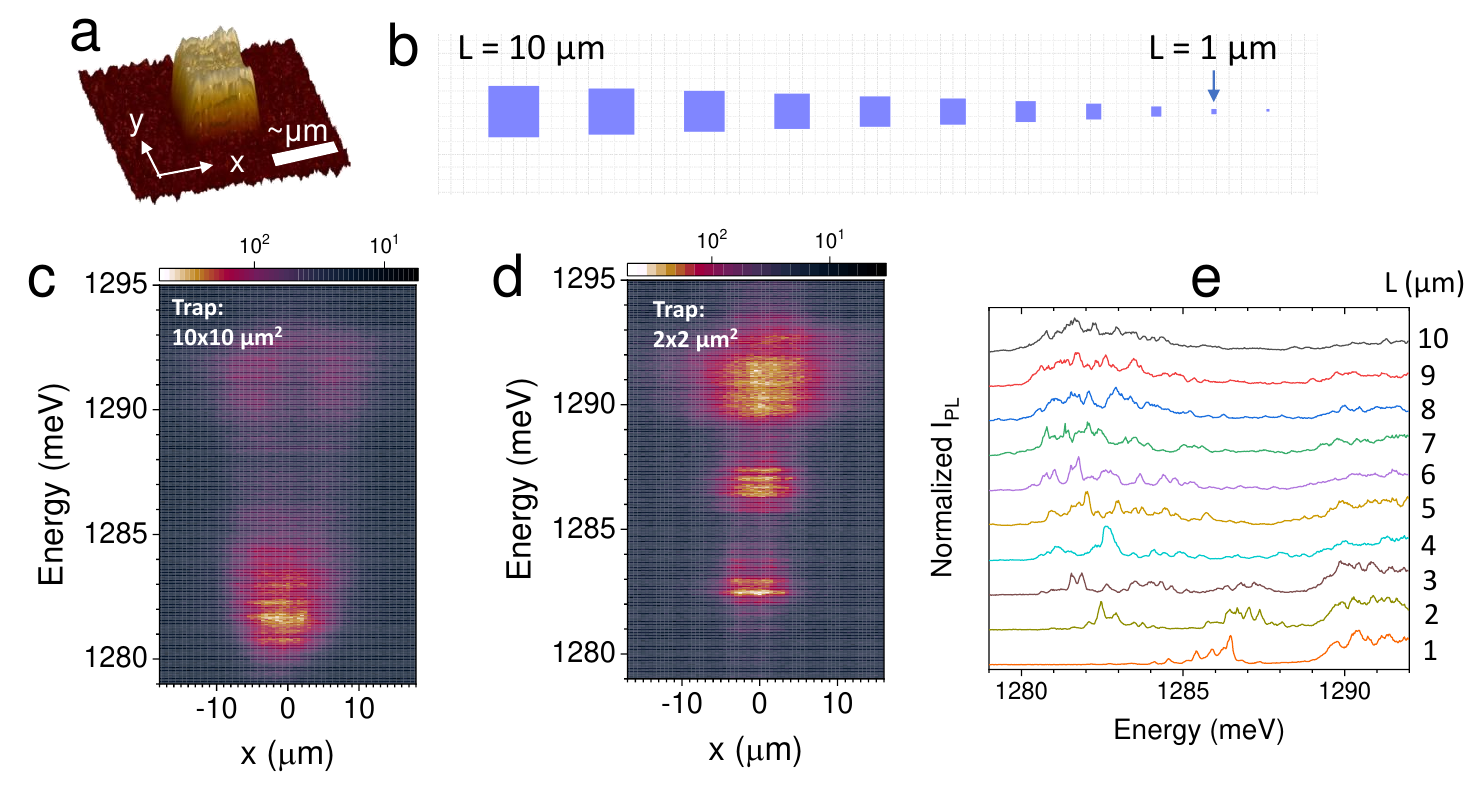}
	\caption{
        {\bf Photonic confinement.}
        {\bf a} Exemplary AFM of a $L \times L = 4 \times 4 ~ \mu m^2$ trap. The height of the mesa is $\approx 10$~nm. 
        {\bf b} The size of the traps varies from $L = 10 ~\mu m$ down to $L = 1 ~\mu m$.
        {\bf c,d} Spatially and spectrally resolved photoluminescence of non-resonantly excited $10 \times 10 ~ \mu m^2$ and $2 \times 2~ \mu m^2$ traps, respectively. The color bars represent the PL intensity in arbitrary units.
        {\bf e} Spatially integrated spectra as a function of $L$.
        }
\label{Fig_SI_Confinement}
\end{figure*}

Figure~SI~\ref{Fig_SI_Confinement}a shows an exemplary atomic force microscope image of a $5 \times 5 \mu \mathrm{m^2}$ trap after the overgrowth. Optical lithography enables the fabrication of traps with lateral dimension ($L$) down to $L = 1 \mu \mathrm{m}$. We have investigated confinement in traps with $L = 10 \mu \mathrm{m}$ down to $L = 1 \mu \mathrm{m}$, cf. Figure~SI~\ref{Fig_SI_Confinement}b.

Figures~SI~\ref{Fig_SI_Confinement}c,d depict spectrally and spatially resolved photoluminescence (PL) maps of $L = 10 \mu \mathrm{m}$ and $L = 2 \mu \mathrm{m}$ traps under weak non-resonant optical excitation at 5~K. As discussed in the main text, above approximately 1290~meV, the emission arises from the spatially extended continuum of states from the barrier (which is the spatially extended etched region). The PL below 1290~meV originates from the confined states of the trap and is localized in space. In the case of the $L = 2 \mu \mathrm{m}$ trap, the PL is also structured in energy, forming well-defined confined modes. The PL consists of narrow lines, which correspond to quantum dot resonances. Spatially integrated spectra of traps with different $L$ are summarized in Fig.~SI~\ref{Fig_SI_Confinement}e. As expected, the lowest confined energy blueshifts with decreasing $L$. For traps with $L < 5 \mu \mathrm{m}$, individual confined modes can be identified.

\newpage
\section{Trap dispersion}
\label{appendix:reference}

Here, we discuss the angular distribution of the PL of QDs within traps (discussed in the section above). The measurements were done using a standard back focal plane imaging technique, which allows to obtain the angular distribution of the emitted light, see e.g., Ref.~\cite{kuznetsov2018Phys.Rev.B}. For MCs, this angle is directly related to the in-plain momentum of photons. 

\begin{figure*}[h]
	\centering
	\includegraphics[width=0.7\textwidth, keepaspectratio=true]{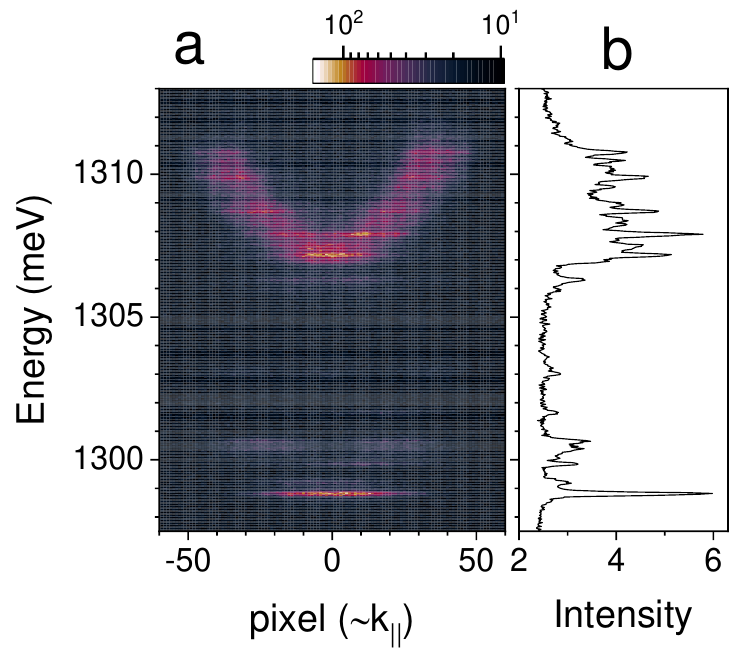}
	\caption{
        {\bf Angular dispersion of trap emission.}
        {\bf a} Emission of $4 \times 4 ~ \mu m^2$ trap resolved in energy and in-plain momentum, $k_{||}$, prop. to the emission angle. The color bar represents the PL intensity in arbitrary units.
        {\bf b} Spectrum integrated over $k_{||}$. The intensity is expressed in arbitrary units.
        }
\label{Fig_SI_TrapMomentum}
\end{figure*}

Figure~SI~\ref{Fig_SI_TrapMomentum}a shows an angle-resolved PL of a $4 \times 4 \mu \mathrm{m^2}$ trap. The emission with parabolic dispersion above 1306~meV originates from the spatially extended etched region of MC, where there is no lateral confinement. An angle-integrated spectrum is shown in  Fig.~SI~\ref{Fig_SI_TrapMomentum}b. It shows that the emission consists of overlapping spectrally narrow lines. Hence, we directly see that the emission of QDs is filtered by the parabolic dispersion of the cavity. Below 1306~meV, PL spectrum is quantized in energy, leading to the flat (dispersion-less) emission. From the dispersive part, we can determine the optical linewidth of about 0.9~meV for the zero in-plane momentum and, thus, the quality factor $Q_\mathrm{o} \approx 1450$.

\newpage
\section{Trap PL excitation power dependence}
\label{appendix:reference}

\begin{figure*}[h]
	\centering
	\includegraphics[width=0.7\textwidth, keepaspectratio=true]{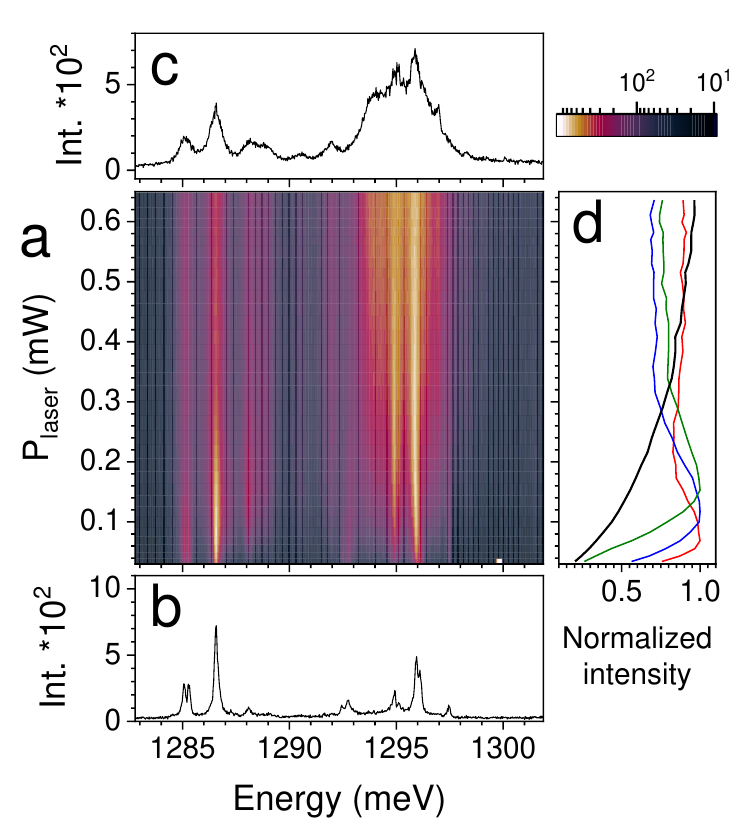}
	\caption{
        {\bf Emission of QDs in a trap vs. excitation power.}
        {\bf a} Spatially integrated PL spectrum of a $4 \times 4 ~ \mu \mathrm{m}^2$ trap as a function of the non-resonant laser power ($P_\mathrm{laser}$) at 5~K. The color bar represents the PL intensity in arbitrary units.
        Exemplary spectra fo the lowest and the highest $P_\mathrm{laser}$ are shown ({\bf b}) below and ({\bf c}) above, respectively. The intensity is expressed in arbitrary units.
        {\bf d} Energy-integrated normalized intensity profiles vs. $P_\mathrm{laser}$ for the ground state (red curve), first (blue curve) and second (green curve) excited state and the continuum (black curve).
        }
\label{Fig_SI_TrapExcPower}
\end{figure*}

At low optical excitation powers, the trap emission consists of spectrally narrow lines due to the QD resonances. Figure~SI~\ref{Fig_SI_TrapExcPower}a shows a PL map of a $4 \times 4 ~ \mu \mathrm{m}^2$ trap as a function of the power of the non-resonant laser ($P_\mathrm{laser}$). At the lowest power $P_\mathrm{laser} = 60 \mu \mathrm{W}$, the spectrum consists of narrow lines, cf. Fig.~SI~\ref{Fig_SI_TrapExcPower}b. As the $P_\mathrm{laser}$ increases, the intensity increases and the lines broaden spectrally. The spectrum for the highest $P_\mathrm{laser} = 0.64$~mW in Fig.~SI~\ref{Fig_SI_TrapExcPower}c shows only the spectral envelopes given by the spectral width of the confined modes of the trap. Here, we use the spectral width of the PL under this condition as another way to determine the optical quality factor of the confined modes: $Q_\mathrm{o} \approx 1600$, which is comparable to the one determined from the PL dispersion in the previous section. 

We note that the PL of the photonic continuum (above 1293 meV) shows the typical saturation of the integrated PL as a function of $P_\mathrm{laser}$, cf. black curve of Fig.~SI~\ref{Fig_SI_TrapExcPower}d. However, the integrated PL of the individual confined levels shows a more complex behavior with a fast initial growth of the intensity, followed by a partial decrease and then becoming constant, as depicted by the colored curves of Fig.~SI~\ref{Fig_SI_TrapExcPower}d. This behavior is currently under investigation.

\newpage
\section{Bulk acoustic resonators}
\label{appendix:reference}

\begin{figure*}[!h]
	\centering
	\includegraphics[width=1\textwidth, keepaspectratio=true]{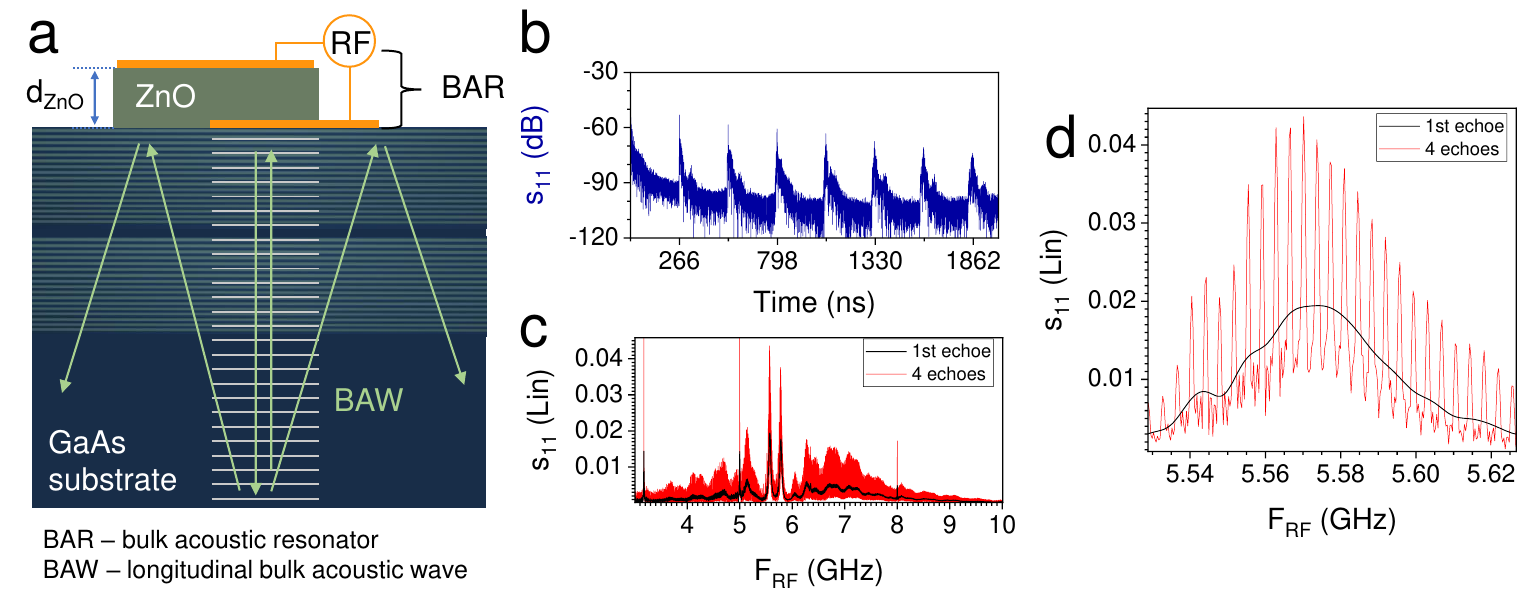}
	\caption{
        {\bf Bulk acoustic wave resonators (BAR).}
        {\bf a} Sketch of a BAR on top of a planar MC. When driven by the radio frequency signal, BAR generates a bulk acoustic wave (BAW, horizontal thin lines). The BAW undergoes multiple reflections at the top and bottom surfaces (vertical arrows), with partial lateral propagation (tilted arrows).
        {\bf b} Time-domain $s_{11}$ response of a BAR in the 3--10~GHz range. Equidistant peaks correspond to BAW echoes at 5~K.
        {\bf c} Time-gated $s_{11}$ obtained for one and four echoes of the panel $b$.
        {\bf d} An expanded range of the curves in $c$ around the cavity mode.
        }
\label{Fig_SI_BAR}
\end{figure*}

Bulk acoustic wave resonators (BAR) were fabricated on the top of the MC sample using a procedure described in Ref.~\cite{machado2019Phys.Rev.Appl.}. A BAR is schematically shown in Fig. SI~\ref{Fig_SI_BAR}a. It consists of a bottom Ti/Au (10/30 nm) contact, a layer of piezoelectric ZnO ($d_\mathrm{ZnO} \approx 300$~nm) and a top Ti/Al/Ti (10/30/10) contact. A BAR is an acoustic resonator. To a first approximation, its central frequency is $F_\mathrm{BAR} \approx v_\mathrm{ZnO} / (2 d_\mathrm{ZnO})$, where $v_\mathrm{ZnO} = 6070$~m/s is the velocity of a longitudinal bulk acoustic wave (BAW). In reality, due to the mass-density of the contacts the frequency is lower. Since BAR is a low-Q acoustic resonator, it has a bandwidth of several GHz around $F_\mathrm{BAR}$~\cite{machado2019Phys.Rev.Appl.}. When a monochromatic microwave signal (RF) with a frequency $F_\mathrm{RF}$ around $F_\mathrm{BAR}$ is applied to the BAR, the latter generates a BAW of the same frequency ($F_\mathrm{RF}$). Reciprocally, a BAR can convert BAWs into an RF signal.

This bi-directional conversion between RF and BAWs can be used to study the generation and propagation of GHz BAWs in complex structures such as MCs~\cite{kuznetsov2021Phys.Rev.X} using conventional microwave instruments, e.g. a vector network analyzer (VNA). In this method, VNA sweeps the frequency and measures the amplitude and the phase of the reflected RF signal corresponding to the $s_{11}$ parameter. Since both the phase and the amplitude are available, one can obtain an effective time response of the signal via the inverse Fourier transform. This allows one to differentiate between the electromagnetic and acoustic contributions due to the huge difference in their propagation velocities.

Figure SI~\ref{Fig_SI_BAR}b shows a time-domain representation of the $s_{11}$ of a BAR on a MC with quantum dots over the 3--10~GHz range recorded at 5~K. One can clearly see time-periodic peaks corresponding to the consecutive BAW echoes due to the multiple reflections at the bottom and the top sides of the sample. The time-separation between the adjacent echoes corresponding to the round trip time, is $\Delta T_\mathrm{rt} = 2d_\mathrm{sample} / v_\mathrm{ZnO}$, where the total thickness of the sample is the sum of the MC and substrate thicknesses $d_\mathrm{sample} = d_\mathrm{MC} + d_\mathrm{subst.}$. In the present case $d_\mathrm{subst.} = 350 \pm 25 \mu$m. 

Figure SI~\ref{Fig_SI_BAR}c shows the frequency components of the first echo (black curve) as well as of the first four echoes (red curve). This representation corresponds to an effective transmission measurement through the sample. We see that the response covers a broad range of frequencies from 3~GHz to above 10~GHz. The complex spectral shape of the curves originates from the fact that the MC acts as a filter for some acoustic frequencies. Overall, there is a correspondence between the experiment and the simulation of Fig. SI~\ref{Fig_SI_MC-Simulations}c. The region of small amplitudes between 5.5~GHz and 6~GHz is the acoustic stopband of the MC. There are two narrow peaks within the stopband. The one at about 5.6~GHz is likely the intended acoustic cavity mode. The other around 5.8~GHz is possibly the mode formed by the BAW and the upper mirror, as has already been observed in Ref.~\cite{kuznetsov2021Phys.Rev.X}. The interference of the acoustic echoes leads to formation of high-Q Fabry-Perot modes, cf. Fig. SI~\ref{Fig_SI_BAR}d. Finally, acoustic echoes propagate laterally, as is schematically shown in Fig. SI~\ref{Fig_SI_BAR}a by the solid green arrows, which allows us to use BARs with the apertures for the optical experiments~\cite{kuznetsov2021Phys.Rev.X}.

\newpage
\section{Raw PL vs. $A_\mathrm{BAW}$ data and filtering procedure}
\label{appendix:reference}

\begin{figure*}[!h]
	\centering
	\includegraphics[width=0.7\textwidth, keepaspectratio=true]{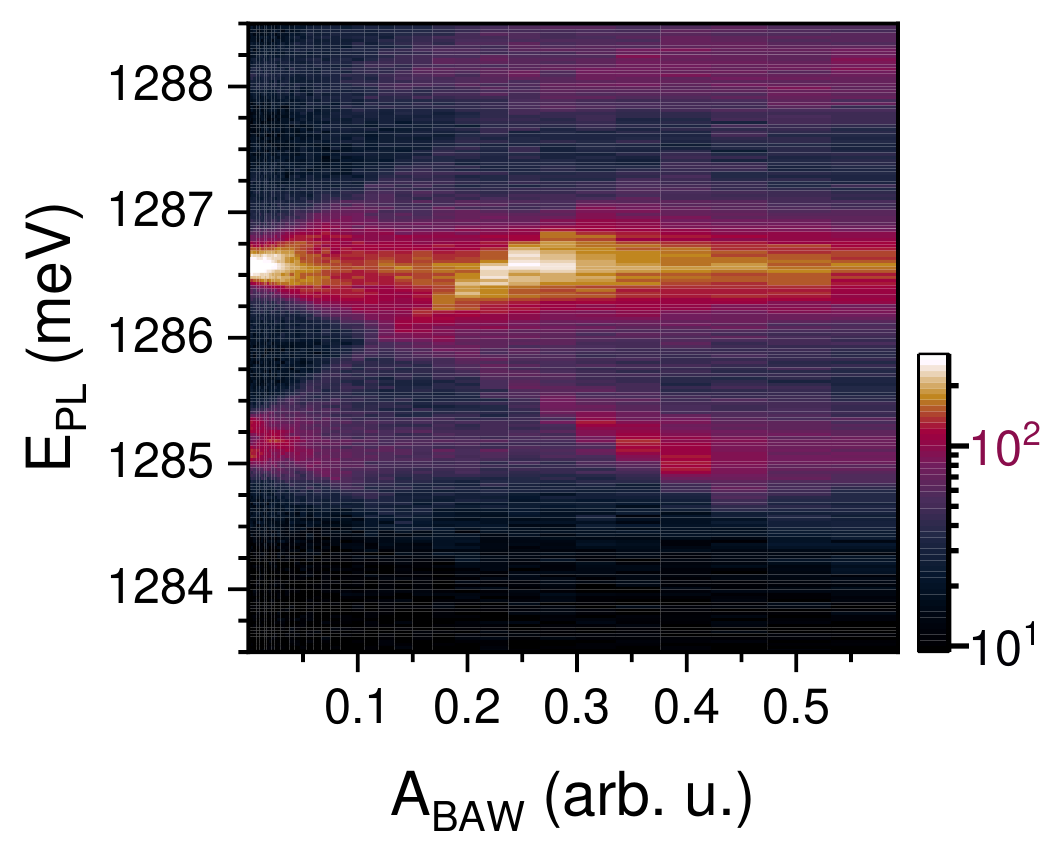}
	\caption{
        {\bf As-measured data.}
        A PL map with as-measured data, which was used to obtain a filtered image shown in the Fig.~3b of the main text. The color bar represents the PL intensity in arbitrary units.
        }
\label{Fig_SI_RAW_PL}
\end{figure*}

Figure SI~\ref{Fig_SI_RAW_PL} shows a PL map of an as-measured QD spectrum in a trap as a function of the modulation amplitude. This data was post-processed to obtain Fig.~3b of the main text. A Wolfram Mathematica software was used to apply a bandpass filter to the image data was to remove low frequency PL components along the y-direction (energy axis). Specifically, low-frequency filter cut-off was set to 0.17, while the kern was set to 25.

\newpage
\section{Surface roughness after MBE-overgrowth}
\label{appendix:reference}

\begin{figure*}[!h]
    	\centering
    		\includegraphics[width=0.8\textwidth, keepaspectratio=true]{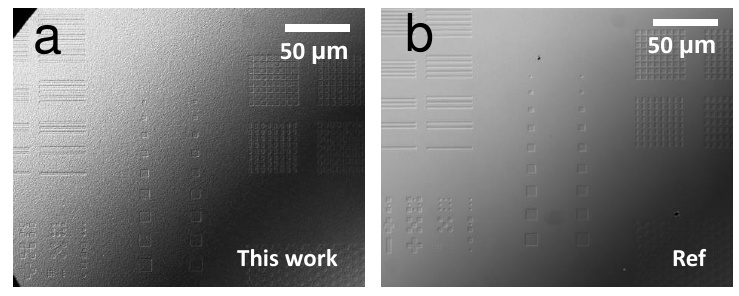}
    		\caption{
               Optical images (50x magnification) of the surface of two microcavities after the MBE-overgrowth comparing optical surface roughness of a MC in this work (on the left) and of a MC with a similar design and patterning from our recent work~\cite{kuznetsov202a6Nat.Photon.} (on the right). Various regular shapes (lines, squares and circles) visible in both images, are traps defined in the MC spacer. Both images were taken under nominally identical conditions.   
            }
    	\label{Fig_SI_Surface_roughness}
    \end{figure*}

    As we mention in the text, sometimes the MBE-overgrowth after patterning of the mesas in the cavity spacer leads to an increased interfacial roughness, which is detrimental for the optical quality of the microcavity (MC). In the Fig.~\ref{Fig_SI_Surface_roughness}, we show two optical microscope images (50x magnification) of the sample used in this study (left image) and an exciton-polariton MC (right image) with nearly identical structure. The overgrown QD-MC has a rough surface, while polariton MC has optically flat surface, and we determined $ { Q \approx 5000 }$ at 10 K~\cite{kuznetsov202a6Nat.Photon.}. Importantly, we are using material platform, where for very similar MCs, Q-factors exceeding 10000 (linewidth of approx. 0.1 meV) are routinely obtained.

\newpage
\section{Photonic confinement simulations}
\label{appendix:reference}

Here we examine spectral and spatial photonic modes in an intracavity trap defined by etching of the cavity spacer~\cite{kuznetsov2018Phys.Rev.B} using simulations. 
Trap spectra have been calculated following the procedure described in the Ref.~\cite{kuznetsov2018Phys.Rev.B}. In short, we determined photonic energy levels by numerically solving the Schrödinger equation for a potential that is given by a graded interfacial potential (which approximates the smooth walls of the mesa in the spacer). Originally, the procedure was used to calculate exciton-polariton spectrum in a mesa. In the present case, only photonic modes inside and outside of the mesa have been considered.

\begin{figure*}[!h]
	\centering
	\includegraphics[width=0.7\textwidth, keepaspectratio=true]{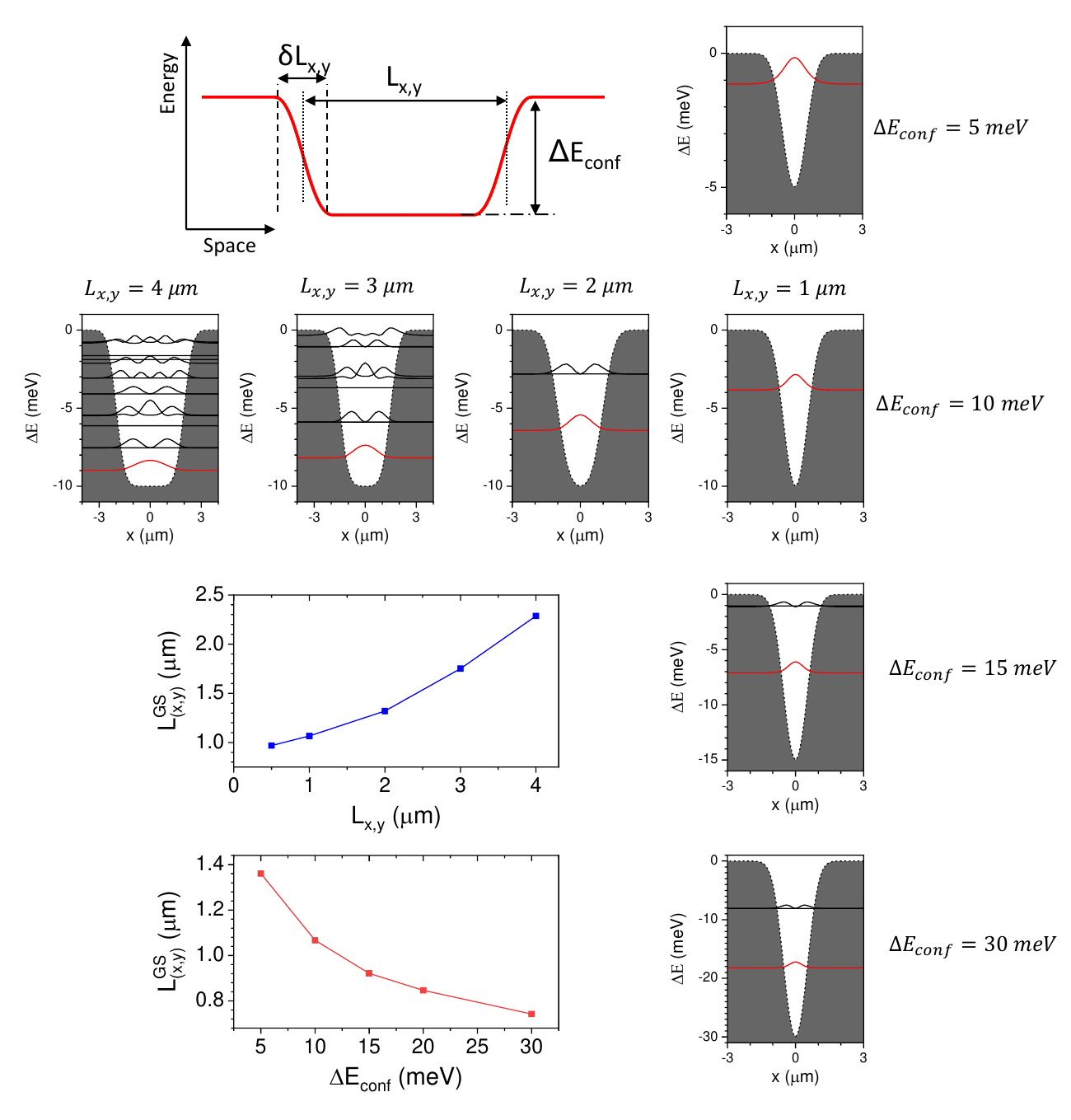}
	\caption{
        {\bf Simulation of photonic trap spectrum vs. trap size and confinement depth.}
        Square trap spatial potential cross-section is approximated by the red curve, where $L_{x,y}$ is the nominal size, $\delta L_{x,y}$ is the sidewall width and $\Delta E_{conf}$ is the confinement energy. $\Delta E$-$x$ plots show calculated spectra, where the dashed lines show the confinement potential and solid lines are squared photonic wavefunctions. Red curves in spatial-energy plots designate the photonic ground state.   
        }
\label{Fig_SI_Photonic_confinement_simulations}
\end{figure*}

The main results are summarized in Fig.~SI~\ref{Fig_SI_Photonic_confinement_simulations}, which shows a dependence of the square trap (with smooth sidewalls) spectrum on nominal trap size ($L_{x,y}$) for fixed confinement potential ($\Delta E_{conf}$), horizontal row, and on the confinement potential for a fixed trap size $L_{x,y} = 1~\mu$m. In the simulation, the trap sidewalls are approximated by a smooth function with width $\delta L_{x,y} = 0.8 \mu$m. The solid lines in the plots represent squared photonic wavefunctions.

One can see that as the $L_{x,y}$ is reduced from 4 to 1 $\mu$m, the number of confined modes reduces, their separation increases and the energy of the fundamental mode (the ground state, GS) is shifted up in energy. The smallest simulated trap with $L_{x,y} = 1~\mu$m fits only a single mode. As the potential depth, designated by $\Delta E_{conf}$, is increased (in practice this can be done by increasing the spacer etching depth), the GS lowers in energy and becomes more localized (smaller spatial extent).

From the simulations we determined the spatial extent of the GS ($L_{x,y}^{GS}$), which we defined as the 1/e width the GS squared wavefunction. Dependencies of $L_{x,y}^{GS}$ on $L_{x,y}$ and $\Delta E_{conf}$ are shown by the plots in the figure. One can see that $L_{x,y}^{GS} \leq 1~\mu$m can be reached for $L_{x,y} < 1~\mu$m and $\Delta E_{conf} > 10$~meV.

\newpage
\section{Purcell factors}
\label{appendix:reference}

    Quality factors and Purcell enhancement of very similar MCs with traps (mesas) lithographically defined in the spacer have been theoretically analyzed in Ref.~\cite{sultanivala2025J.Phys.Photonics} for different mesa diameters, mesa thickness and optical top-DBR pair number $N_ {DBR}$. For mesas with parameters similar to our work, e.g., diameters of $1 ~\mu$m to $3 ~\mu$m, mesa height of 13 nm, $N_ {DBR} = 33$, Q-factors up to 60000 and $ {F_{P}}$ up to 70 have been predicted~\cite{sultanivala2025J.Phys.Photonics}. Although, there is a non-trivial dependence of the Q and $ {F_{P}}$ on the mesa diameter.
    
    To estimate on-resonance Purcell factor, given by $ {F_{P} = 3 Q_{C} / (4 \pi^2 V_{C}) * (\lambda_{C}/n_{C})^3}$, for an optical wavelength $ {\lambda_{C}}$, we need to know the mode volume of a MC photonic mode $ {V_{C}}$ and quality factor $ {Q_{C}}$. For the current MC design, we calculated $ {Q_{C}} = 20000$ from the simulated reflectivity spectrum, Fig.~SI~1d. 
    
    The photonic mode is localized in a small trap. Therefore, the value of  $ {V_{C}}$ depends on its extent into the DBR as well as on the transverse size of the mode defined by the trap. For a gaussian ground state (GS) mode of the trap, we write  $ {V_{C} = L_{eff} * A_{GS}}$. Here, $ {L_{eff} = L_{spacer} + 2 L_{pen}}$, where $ {L_{spacer}}$ is the spacer thickness and $ {L_{pen} = \lambda_{C} / (4 \Delta n)}$ is the approximate 1/e penetration depth of the optical intensity of the mode into the DBR at normal incidence, which depends on the refractive index contrast of the DBR pairs ($\Delta n$)~\cite{panzarini1999Phys.SolidState}. The value of the GS, the area is $ {A_{GS} = \pi L_{x} L_{x} / 4}$, where $ { L_{x,y}^{GS} }$ is the 1/e width of the Gaussian mode. We used simulations of the lateral optical confinement described in SI-Sec. 9 to determine $ {A_{GS}}$. For the current MC design, we obtain  $ {L_{eff} = 1.2~\mu m}$. The table~\ref{table_TrapsCalculated} below shows (ideal) values of $ {F_{P}}$ for different trap sizes. We note that, for $ {1 \times 1~\mu m^2}$ and  $ {2 \times 2~\mu m^2}$ traps, our values are moderately close the ones obtained in Ref.~\cite{sultanivala2025J.Phys.Photonics} for similar cavity and mesa parameters. The table shows the corresponding modified spontaneous emission rates for a QD $ { \Gamma_{SE}^{cav} = 1 / \tau_{SE}^{cav} }$, where $ { \tau_{SE}^{cav} = \tau_{SE}^{bare} / F_{P} }$ and the unmodified radiative lifetime is $ { \tau_{SE}^{bare} = 1000~ps}$.

    \begin{table}[h]
      \centering
    \begin{threeparttable}
      \caption{Calculated values of x,y dimensions of the GS mode for several traps for potential depth of 10 meV, and the corresponding values of the Purcell enhancement and modified spontaneous emission rates (assuming $ {\tau_{SE}^{bare} = 1000 ~ ps}$). }
      \begin{tabular}{|| c | c  | c | c  | c   | c  || }
     \hline
     \hline
     Nominal size   & $ {L_{x,y}^{GS}}$   &   $ {A_{GS} (\mu m^2)}$ & $ {F_{P}}$ & $ {\Gamma_{SE}^{cav} (GHz)}$ \\
     \hline
     $ {4 \times 4 ~ \mu m^2}$ &  2.3   &  4.1  &  8  &  8  \\
     $ {3 \times 3 ~ \mu m^2}$ &  1.75    &  2.4  &  14  &  14  \\
     $ {2 \times 2 ~ \mu m^2}$ &  1.3    &  1.4  &  25  &  25  \\
     $ {1 \times 1 ~ \mu m^2}$ &  1.07   &  0.9  &  37  &  37  \\
    \hline    
    \hline
    \end{tabular}
      \begin{tablenotes}
      \item[]
      \end{tablenotes}
    \label{table_TrapsCalculated}
    \end{threeparttable}
    \end{table}

\newpage
\section{Time-evolution of exciton population under modulation}
\label{appendix:reference}

\begin{figure*}[!h]
	\centering
	\includegraphics[width=0.75\textwidth, keepaspectratio=true]{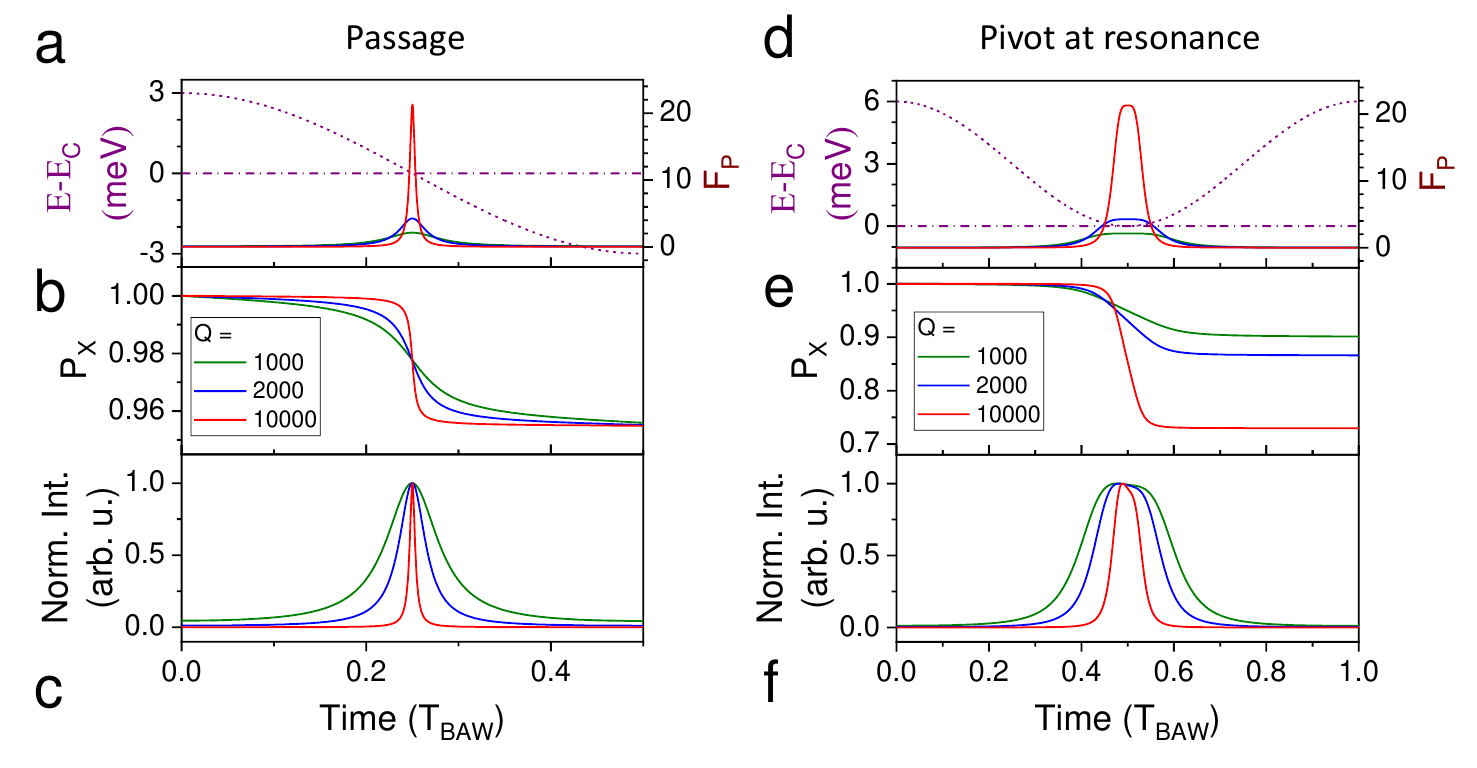}
	\caption{
        {\bf Simulation of the passage vs. pivot on resonance modulation regimes.}
        {\bf a, d} Energy vs. time for the exciton (dashed lines) and photon (dash-dotted lines) and Purcell factor ($F_{P}$, solid lines). The energy scale is relative to the photon energy.  
        {\bf b, e} Evolution of the QD exciton population ($P_{X}$).
        {\bf c, f} Evolution of the normalized intensity.
        The time scale is in units of the modulation period $ { T_{BAW} }$. Colored solid curves correspond to different Q-factors indicated in the panels {\bf b \& e}. 
        }
\label{Fig_SI_Resonant_vs_Passage_modulation}
\end{figure*}

Here, we show simulated time-evolution of exciton population ($P_{X}$) and intensity for two different modulation scenarios: (i) passage of the exciton through the cavity mode (Fig.~SI~\ref{Fig_SI_Resonant_vs_Passage_modulation}a) and (ii) pivot of the exciton at resonance with the cavity mode (Fig.~SI~\ref{Fig_SI_Resonant_vs_Passage_modulation}d). In each case, calculations were done for three values of cavity quality factor ($Q$) and fixed modulation frequency $F_{BAW} = 5$~GHz. The calculation procedure is describe in the main text. 

In the first scenario, the exciton rapidly passes through the cavity mode, leading to a sharp decrease of the exciton population (Fig.~SI~\ref{Fig_SI_Resonant_vs_Passage_modulation}b) and emission of a symmetric pulse of light (Fig.~SI~\ref{Fig_SI_Resonant_vs_Passage_modulation}c). The temporal width of the pulse decreases with the increasing Q-factor and for $Q = 10000$ reaches $\tau_{SP} \approx 0.01 \times T_{BAW}$, where $T_{BAW} = 1/F_{BAW}$. For $F_{BAW} = 5$~GHz we obtain $\tau_{SP} \approx 2$~ps. We note that the total change in the exciton population accumulated over the passage ($\Delta P_{X}$) is independent of $Q$, while only the temporal duration of the $P_{X}$ step changes. This is in agreement with Ref.~\cite{peinke2021LightSci.Appl.}. 

The second scenario has a drastically different behavior. Here, the value of $\Delta P_{X}$ increases with increasing $Q$, cf. Fig.~SI~\ref{Fig_SI_Resonant_vs_Passage_modulation}e. This scenario is discussed in detail in the main text. We point out that the trade-off is the increased SP pulse duration relative to $T_{BAW}$ for the scenario (ii) as compared to the scenario (i) for the same $Q$, cf. Fig.~SI~\ref{Fig_SI_Resonant_vs_Passage_modulation}f.

\newpage
\section{Raw $g^{(2)}$ data for QD in a trap}
\label{appendix:reference}

\begin{figure*}[!h]
	\centering
	\includegraphics[width=0.75\textwidth, keepaspectratio=true]{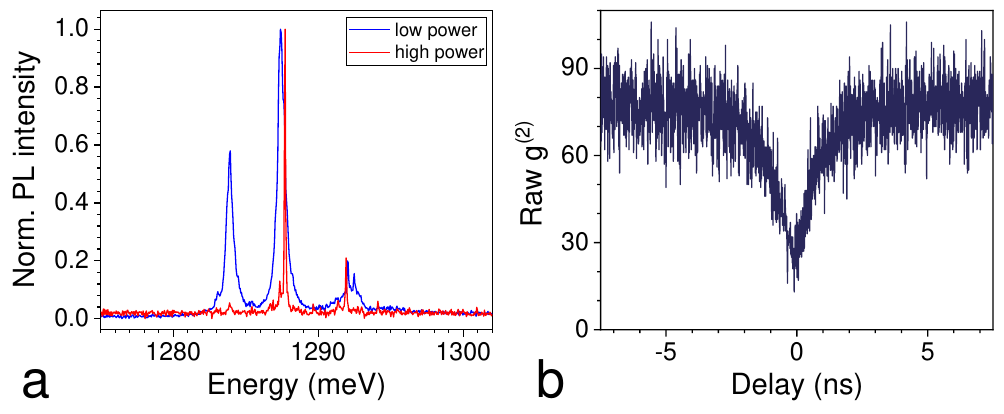}
	\caption{
        {\bf Raw $g^{(2)}$ data.}
        {\bf a} PL spectra of a trap under low- and high-power optical excitation.
        {\bf b} Raw $g^{(2)}$ as a function of time delay.
        }
\label{Fig_SI_Raw g(2)}
\end{figure*}

Figure~SI~\ref{Fig_SI_Raw g(2)}a shows normalized PL spectra of a trap used for $g(2)$ measurements presented in the main text under low and high optical excitation. Under low excitation three very narrow resonances can be seen, corresponding to possibly multiple QDs within the same trap. For the $g(2)$ measurement, emission of the strongest peak was selected using a bandpass filter with approx. 2 nm bandwidth. Figure~SI~\ref{Fig_SI_Raw g(2)}b shows raw $g^{(2)}$ values as a function of delay. The normalized $g^{(2)}$ in the main text, was obtained by dividing the raw $g^{(2)}$ values by the average $g^{(2)}_{long} = 80$ at long delays > 7 ns.

\newpage
\vskip 1 cm
\bibliography{ZoteroBib}

